\documentclass[letterpaper,11pt]{article}
\usepackage{graphicx}
\usepackage[margin=1in]{geometry}
\usepackage{microtype} %if unwanted, comment out or use option "draft"
\usepackage{amssymb}
\usepackage{amsmath}
\usepackage{amsthm}
\usepackage{bm}
\usepackage{xspace}
\usepackage{epsfig}
\usepackage{ifpdf}
\usepackage{url}
\usepackage[hidelinks]{hyperref}
\usepackage{latexsym}
\usepackage{color}
\usepackage{makeidx}
\usepackage{stackrel}
\usepackage{amscd}
\usepackage[all]{xy}
\usepackage{lineno}
\usepackage{algorithm}
\usepackage[noend]{algpseudocode}
\algnewcommand{\LineComment}[1]{\Statex\hspace{\algorithmicindent}\(\triangleright\) #1}
\algnewcommand\algorithmicforeach{\textbf{for each}}
\algdef{S}[FOR]{ForEach}[1]{\algorithmicforeach\ #1\ \algorithmicdo}
\usepackage{varwidth}
\usepackage{framed}
\usepackage{pb-diagram}
\usepackage[dvipsnames]{xcolor}
\usepackage[labelformat=parens]{subfig}
\usepackage[symbol]{footmisc}
\usepackage{mathtools}
\usepackage{etoolbox} %% for the line spacing of algpseudocode
\usepackage[normalem]{ulem}
\usepackage{stmaryrd}
\usepackage[scr=boondoxo]{mathalfa}
\usepackage{soul}
\usepackage{authblk}
\usepackage{multirow}
\usepackage[above,below]{placeins}
\usepackage{makecell}
\usepackage{booktabs} %% for better tabular

\makeatletter
\expandafter\patchcmd\csname\string\algorithmic\endcsname{\itemsep\z@}{\itemsep=0.25ex}{}{}
\makeatother

\usepackage{enumitem}
\setlist[itemize]{leftmargin=\parindent}
\setlist[enumerate]{leftmargin=\parindent}
\setlist[description]{font=\bfseries,leftmargin=\parindent}

\theoremstyle{plain}

    \newtheorem*{property*}{Property}
\theoremstyle{definition}

\newcommand{\Real}{\mathbb{R}}

\newcommand{\given}{\,|\,}
\newcommand{\Set}[1]{\{#1\}}

\newcommand{\specsize}{\mu}
\newcommand{\perslen}{\lambda}

\newcommand{\Bcal}{\mathcal{B}}

\newcommand{\brm}{\text{b}}

\newcommand{\Csf}{\mathsf{C}}
\newcommand{\Gsf}{\mathsf{G}}

\newcommand{\aG}{\alpha}

\newcommand{\OG}{\Omega}

\makeatletter
\newcommand{\algmargin}{\the\ALG@thistlm}
\makeatother
\newlength{\whilewidth}
\algdef{SE}[parWHILE]{parWhile}{EndparWhile}[1]
  {\parbox[t]{\dimexpr\linewidth-\algmargin}{%
     \hangindent\whilewidth\strut\algorithmicwhile\ #1\ \algorithmicdo\strut}}{\algorithmicend\ \algorithmicwhile}%
\algdef{SE}[parIF]{parIf}{EndparIf}[1]
  {\parbox[t]{\dimexpr\linewidth-\algmargin}{%
     \hangindent\whilewidth\strut\algorithmicif\ #1\ \algorithmicthen\strut}}{\algorithmicend\ \algorithmicif}%
\algnewcommand{\parState}[1]{\State%
  \parbox[t]{\dimexpr\linewidth-\algmargin}{\strut #1\strut}}

\begin{document}

\title{Topological Filtering for 3D Microstructure Segmentation}
%3D Microstructure Segmentation by Topological Filtering}
%\title{3D Micro-structure Image Segmentation aided by Persistent Homology:
%Mapping Tortuosity in 3D Porous Structures}

% \author{Tamal K. Dey\thanks{\red Department of Computer Science and Engineering, The Ohio State University. \texttt{dey.8@osu.edu}}
% \and Tao Hou\thanks{\red Department of Computer Science and Engineering, The Ohio State University. \texttt{hou.332@osu.edu}}
% }

\newlength{\authorspace}
\setlength{\authorspace}{2em}

% \author{\red(email, affiliation)
% Anand V. Patel\hspace{\authorspace} Tao Hou\hspace{\authorspace} Juan D. Beltran\hspace{\authorspace} 
% \\ Tamal K. Dey\hspace{\authorspace} Dunbar P. Birnie, III
% \\
% {\footnotesize 
% Department of Materials Science and Engineering, Rutgers University.
% \texttt{dunbar.birnie@rutgers.edu}
% }\\
% {\footnotesize 
% Department of Materials Science and Engineering, Rutgers University.
% \texttt{avp74@soe.rutgers.edu}
% }\\
% {\footnotesize 
% Department of Electrical and Computer Engineering, Rutgers University.
% \texttt{juan.beltran@rutgers.edu}
% }\\
% {\footnotesize 
% Department of Computer Science and Engineering, The Ohio State University.
% \texttt{dey.8,hou.332@osu.edu}
% }\\
%{\footnotesize 
%Department of Computer Science, Purdue University. 
%\texttt{tamaldey,hou145@purdue.edu}
%}
% }

\author[1]{Anand V. Patel \thanks{avp74@soe.rutgers.edu}}
\author[2]{Tao Hou \thanks{hou145@purdue.edu}}
\author[3]{Juan D. Beltran Rodriguez \thanks{jdb334@scarletmail.rutgers.edu}}
\author[2]{\authorcr Tamal K. Dey \thanks{tamaldey@purdue.edu}}
\author[1]{Dunbar P. Birnie, III \thanks{dunbar.birnie@rutgers.edu}}

\affil[1]{Department of Materials Science and Engineering, Rutgers University}
\affil[2]{Department of Computer Science, Purdue University}
\affil[3]{Department of Electrical and Computer Engineering, Rutgers University}

\date{}

\maketitle
\thispagestyle{empty}

% \tableofcontents\vspace{5em}

\begin{abstract}
Tomography is a widely used tool for analyzing microstructures in three dimensions (3D). The analysis, however, faces difficulty because the constituent materials produce similar grey-scale values. Sometimes, this prompts the image segmentation process to assign a pixel/voxel to the wrong phase (active material or pore). Consequently, errors are introduced in the microstructure characteristics calculation. In this work, we develop a filtering algorithm called {\sc PerSplat} based on topological persistence
(a technique used in \emph{topological data analysis})
to improve segmentation quality. 
One problem faced when evaluating filtering algorithms is that real image data in general are not equipped with the `ground truth' for
% information about 
the microstructure characteristics. For this study, we construct synthetic images for which the ground-truth values are known. 
On the synthetic images, we compare the pore \emph{tortuosity} 
%\tamal{`interconnected' goes with what? tortuosity and functional? can we take out `interconnected'?} 
and \emph{Minkowski functionals} (volume and surface area) computed with our {\sc PerSplat} filter and other methods such as total variation (TV) and non-local means (NL-means). 
% Experimental results and visual inspections show that 
Moreover, on a real 3D image,
we visually compare the segmentation results provided by our filter against
%\tamal{`with' was confusing} 
TV and NL-means.
The experimental results indicate that {\sc PerSplat}
provides a significant improvement in segmentation quality.
% reproducing { \anand{binary}} results {\sout{tortuosity}} close to the ground truth. {\sout{, even when the grey-scale values of the phases are similar}}.
%are corrupted with noise were similar\tao{`with noise were similar' does not sound right to me}.  

\end{abstract}

{
  \small	
  \textbf{Keywords: } 
  Tomography, Topological persistence,
  Image segmentation, Image filtering, Tortuosity
}

\newpage
\setcounter{page}{1}

\section{Introduction}

Microstructures are the building blocks of a material: the shape, size, interconnection, and orientation of these microstructures play a key role in defining the ultimate properties and performance. 
For example, in energy storage systems like lithium ion batteries (LIB), these microstructures influence the ability to store energy, the conversion rates, and the diffusion phenomena
%\tao{this part is a little weird to me. need explain.}
\cite{vwoodseth,Tarascon2001,Yamakawa2020,Chen2017,Habte2018,Campos2020,Gireaud2006}.
%{while maintaining the mechanical strength}
Hence, analyzing and visualizing these microstructures help us understand the properties of the materials and design new materials~\cite{Lu}. 
% Such kind of characterization are of two types- 2D analysis and volume average measurements. 

%Optical Microscopy, Scanning Electron Microscopy come under 2D analysis while techniques like X-ray Diffraction, spectroscopies, and X-ray computational tomography (XCT) come under the volume average measurements. Before XCT was introduced, researchers used to use combination of SEM and meticulous etching to create 2D slices and reconstruct a 3D image. Due to lake of 3D visualization by these techniques in material science a new field started emerging in 1960s called computational tomography which is a nondestructive technique. It was Sir Godfrey N. Hounsfield, in 1972 who successfully conducted practical XCT measurements. 
Nowadays, microstructures are frequently examined in 3D using X-ray computed tomography (XCT or CT) \cite{Taiwo2017}. In XCT, X-ray images are collected from many different directions and  are interpreted to yield 3D maps of local X-ray absorption strength quantified by a grey-scale \cite{Wood2018}. 
%are projected on to the sample which exploits the materials ability to absorb X-rays to make 2D grey-scale images. For instance, a dense material will absorb less X-rays as compared to air (low density), this results in a grey-scale image. 
%This image is the map that is interpreted to find the location of particular phases in the 3D space.
%Depending on resolution of the XCT machine used, image acquisition algorithm and sample, the image data will form a histogram with different number of peaks. These peaks correspond to the number of materials that is detected by the machine or in some cases it could be the artifacts (noises). 
%When one plots the voxel (or pixel) values of the images a grey-scale histogram is \tao{mentioning of hisogram here is too abrupt to me, any background?} produced which is normally used to segment the data. 
%could have two or more distinct peaks or one peak which will have another smaller peak/s merged in. 
Besides an extensive usage of XCT in materials science to characterize fuel cell electrodes \cite{Li2019,Ramani2018}, super-capacitor electrodes \cite{Abouelamaiem2018}, porous ceramics \cite{Nickerson2019}, and fiber orientation in composites {\cite{Emerson2017}}, 
XCT is also used in biological applications such as tumor detection \cite{Qaiser2019,Nagarajan2019}, fracture examination~\cite{Nuchtern2015,Grigoryan2003}, and blood clots \cite{Cano-Espinosa2020}.
In battery research, the XCT experiments are performed to measure three important microstructure characteristics, 
i.e., the porosity, the specific surface area, and the tortuosity of pore network \cite{vwoodseth,Nguyen2020,Ebner2014}. 

For the present work, we emphasize microstructural tortuosity as it directly influences current flow rates within the interpenetrating electrolyte phase contained in the pores.
%The porosity tells the ability of battery to store energy and transport energy. The specific surface area tells how much active material is in contact with the electrolyte to understand the efficiency. The tortuosity tells the charging and discharging rate of the battery. 
Typically, before calculating porosity or tortuosity, the data gathered from XCT are processed to assign each voxel/pixel to the solid phase or the pore phase; this process is called {\it image segmentation} (which means {\it binarization} for datasets of this paper). 
% \tao{keep the deleted sentence because we then would be using `binarization' without a proper
% mention all over the paper.}
However, depending on the mixture of phases being studied and their ability to absorb X-rays, a `faithful' binarization of data can be very difficult.
%\sout{image segmentation can be very difficult.
%\tao{and the reason is explained in the following.} Typically, in image segmentation, a `faithful' binarization of data turns out to be one of the most difficult tasks.}
% {\sout{A typical image segmentation process consists of three parts, (i) filtering, (ii) binarization and (iii) post processing. In filtering process, the grey-scale values are altered either by local averages, medians, maxima, or minima.Some of these processes are described in "Related Works". \cite{Pietsch2017} \cite{otsu1979threshold}.}}
%The binarization is the process where the grey-scale is divided into a group of similar values and then the groups are assigned as solid or pore
Hence, an appropriate {\it filtering} before the binarization is often
necessary to improve the quality of binarization and thereby
the accuracy of computed microstructure characteristics~\cite{iassonov2009segmentation,kaestner2008imaging}.

\begin{figure}
  \centering
  \subfloat[]{\includegraphics[width=1\linewidth]{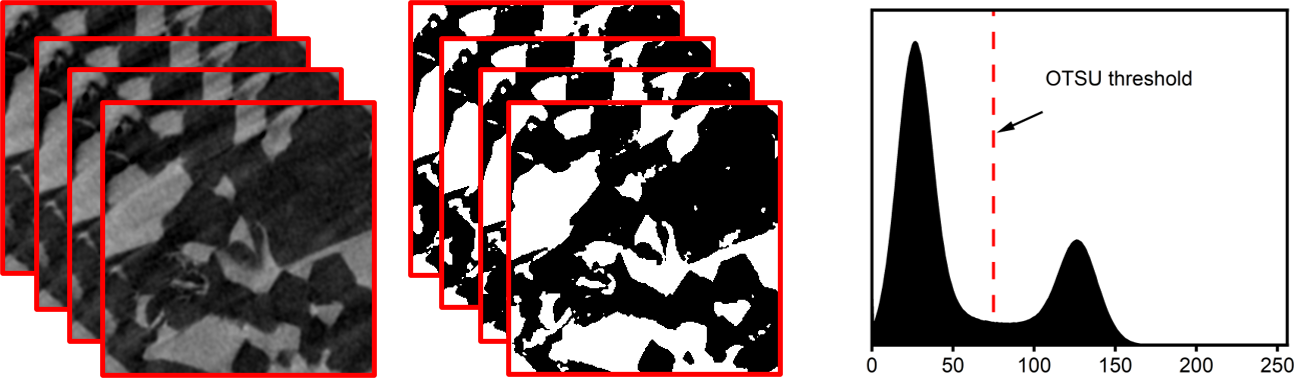}\label{fig:1}}
  \hspace{2em}
  \subfloat[]{\includegraphics[width=1\linewidth]{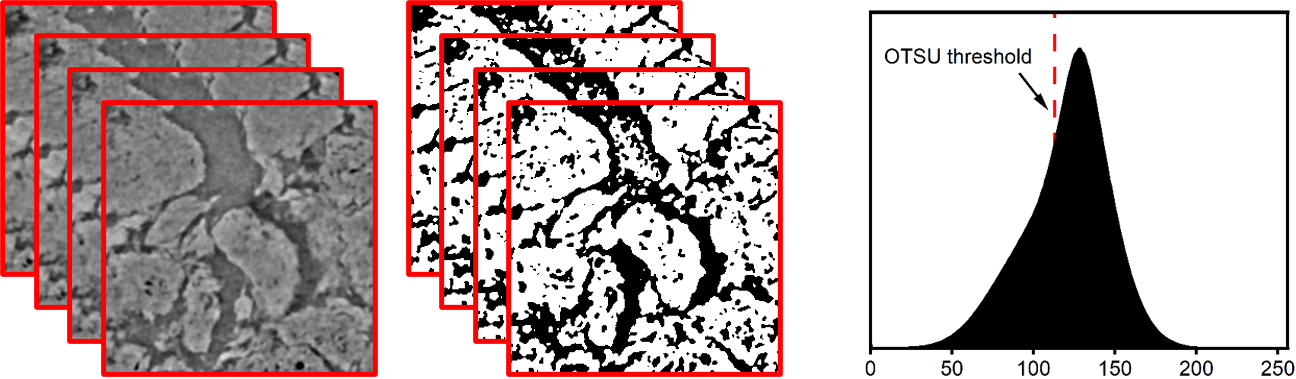}\label{fig:2}}
  % \hspace{1em}
   \caption{\textbf{XCT data.}
   (a) Grey-scale data, binarized data, and histogram of templated porous polymer.
   (b)~Grey-scale data, binarized data, and histogram of commercial electrode.
   }
    \label{fig:results1}
\end{figure}

To illustrate the difficulty faced by a binarization algorithm
without a proper filtering,
we take the popular {\it global thresholding} algorithm 
{\sc Otsu}~\cite{otsu1979threshold} as an example.
% which is a popular method for binarization.
In {\sc Otsu}, a threshold value is picked first,
and then any pixel/voxel above (resp. below) the threshold value is assigned active material (resp. porous region).
% In a binarization process, a threshold value is first picked by an algorithm such as the well-known Otsu's algorithm~\cite{otsu1979threshold} (henceforth called {\sc Otsu}) and then any pixel/voxel above (resp. below) the threshold value is assigned active material (resp. porous region).
%\sout{and any pixel/voxel above that threshold value is assigned phase A (active material) and any pixel/voxel below that threshold value is assigned phase B (porous region)}. 
% {\sc Otsu} algorithm finds a threshold value by maximizing the variance between the two classes (black and white), 
% and therefore it
It is known that
{\sc Otsu} works well for images with {\it bimodal} histograms, 
where peaks of the two classes in the histograms are distinct. 
However, due to its strong reliance on histograms, 
{\sc Otsu} algorithm may have difficulty when the peaks overlap and are not distinguishable.
% , as shown in Figure~\ref{fig:2}. 
%\sout {\tao{this content is also dicussed previously}}. 
%  {\sc Otsu} algorithm also has advantages such as not requiring prior knowledge of the input and having low computational cost. 
For example,
Figure~\ref{fig:1} shows
% the histogram of 
a 3D grey-scale image with a bimodal histogram,
% has two distinct classes 
and the binarization by {\sc Otsu} preserves almost all important regions. 
In contrast,
Figure~\ref{fig:2} shows
% the histogram of 
a 3D image whose histogram is not bimodal,
% has two distinct classes 
and therefore the binarization by {\sc Otsu} does not reflect the true porosity or electrolyte channel.
In addition to the problem described above, 
data collected from XCT also have artifacts, which can potentially clutter the grey-scale values of pixels/voxels~\cite{Pietsch2017, Muller2018}
and hence further hinder the binarization process.

\paragraph{Contribution.}
The motivation of this work is to improve the binarization results of images 
with a {\it new topological filter}.
% whose histograms have overlapping peaks. 
Specifically, we propose a filtering algorithm called {\sc {\sc PerSplat}} based on techniques developed from {\it topological data analysis}~\cite{deycomputational,EH10} (TDA). %, which is a fast-growing field sitting at the intersection of computer theory and Mathematics.
This algorithm exploits topological structures hidden in the data to filter out noise/speckle
so that overlaps in the histogram can be reduced. 
The filtered data fed to a binarization algorithm
% {\sc Otsu} 
afterwards can then have a better binarization result.
Compared to traditional filtering algorithms (such as those described in~\cite{buades2005review,loizou2008despeckle}),
{\sc PerSplat} algorithm has the following advantages:
\begin{enumerate}
    \item 
    Based on the theory of topological persistence~\cite{edelsbrunner2000topological}, 
    {\sc PerSplat} detects {\it global} structures rather than local ones in a multi-scale manner.
    The removed noise can be of any shape which do not need to fit in a rectangular window.
    This is in contrast to traditional despeckling filtering algorithms~\cite{loizou2008despeckle}
    utilizing a moving window, which restricts the shapes
    of detected speckle.
    
    \item 
    In contrast to traditional despeckling or denoising algorithms~\cite{buades2005review,loizou2008despeckle},
    which aim at recovering the ground-truth image,
    our algorithm aims at a filtering of the image so that
    the {\it ensuing binarization} can recover the true segmentation 
    of the phases.
    % \footnote{We also note that {\sc {\sc PerSplat}} only aims at a preprocess of the image so that the later binarization can be more effectively performed.}.
    Achieving such a goal is especially valuable 
    when the ground truth image {\it itself} is hard to binarize.
    As indicated by experiments, 
    % the two peaks of histograms for images can be successfully separated by our algorithm,
    % so that 
    the binarization by a standard scheme (such as {\sc Otsu}) is more reliable
    with {\sc PerSplat}.
    
    \item Some small but significant features can be preserved 
    by inhibiting the removal of those small regions that 
    {\it persist} for a long range of values.
    This is achieved with the help of the well-established
    tool called {\it barcodes}~\cite{EH10,edelsbrunner2000topological} (or {\it persistence diagrams}),
    in which long {\it bars} are considered as significant features
    and short bars are treated as noise.
\end{enumerate}

%\sout{Note that we apply {\sc PerSplat} on 3D data in two ways in this paper: 

%\begin{itemize}
   % \item 
%   \begin{description}
    % \item[{\sc PerSplat-P} (i.e., {\sc PerSplat-}{\it Pixelized}):]
    %{\sc PerSplat-P} (i.e., {\sc PerSplat-}{\it Pixelized}):
    %In this case, we process the grey-scale data slice by slice where the 3D input is treated as a stack of 2D pixelized slices. Each slice is processed with a version of {\sc PerSplat} tailored for 2D input. The outputs for all slices are then stacked to create 3D filtered data.
    
    %\item 
    % \item[{\sc PerSplat-V} (i.e., {\sc PerSplat-}{\it Voxelized}):] 
    %{\sc PerSplat-V} (i.e., {\sc PerSplat-}{\it Voxelized}):
    %In this case, all slices together are treated as a 3D voxelized data which is processed with a version of {\sc PerSplat} tailored for 3D input. The output is also three dimensional grey-scale data.}
    %\tao{I really think this should be defined later in the results section. Otherwise, reader would have to go back and forth to understand the pipeline.}
    % \end{description}
%\sout{In other words, the topological analysis is done simultaneously in all the directions.}
%\end{itemize}

We briefly describe our {\sc PerSplat}
algorithm in Section~\ref{sec:persplat}. In our experiments,
{\sc PerSplat} 
outperforms other general-purpose filtering methods 
(TV~\cite{rudin1992nonlinear,rudin1994total} and NL-means~\cite{buades2005review})
% on both the accuracy of the computed microstructure characteristics (e.g., tortuosity) on 
on both synthetically generated datasets
% and visual inspection on 
and natural datasets; see Section~\ref{sec:results} for details.
% Before presenting the experiments,

%\anand{add few lines in ground truth and synthetic data}\tao{Here maybe a good place to introduce a bit the synthetic data generation.}
%Specifically, we compare the tortuosity values from the output image of {\sc Otsu} alone with those obtained from the output of {\sc {\sc PerSplat}} and {\sc Otsu} together. 

%feature and extent of this algorithm we came up with a methodology where you can check the quality of binarized data by checking the ground truth values of tortuosity. 

\subsection{Related works}

Considering the volume of existing literature on the topic,
we only briefly describe a few image filtering/denoising algorithms
of various types. For a more comprehensive overview,
we recommend the work by Buades el al.~\cite{buades2005review} on denoising for general images,
the work by Loizou and Pattichis~\cite{loizou2008despeckle} on despeckling for ultrasound images,
and the work by Kaestner et al.~\cite{kaestner2008imaging} on image filtering in porous media research.
In our overview, we only describe the algorithms in 2D; their generalization
to 3D is straightforward.
Throughout, 
let $\OG$ denote the domain of the image which is a rectangular region in $\Real^2$,
and  an image is {therefore} treated as a function $v:\OG\to\Real$.

\paragraph{Total variation.}
Total variation (TV) denoising~\cite{rudin1992nonlinear,rudin1994total}
draws upon a
% {\it total variation} (
measurement of regularization
% )
$TV_\OG(u)$ for an image $u$, which is the integral of the gradient magnitude.
% where $Dv$ is a {\it Radon measure}.
% $TV_\OG(v)$ provides
% a measurement of regularization for the image.
Given an input image $v$,
TV denoising minimizes 
$TV_\OG(u)$ for
% the total variation of
the denoised image $u$
subject to the following constraints:
\[\int_\OG(u(x)-v(x))\mathrm{d}x=0\quad\text{and}\quad\int_\OG|u(x)-v(x)|^2\mathrm{d}x=\sigma^2.\]
Alternatively, TV denoising solves a corresponding unconstrained minimization problem:
\[\arg\min_u TV_\OG(u)+\lambda\int_\OG|u(x)-v(x)|^2\mathrm{d}x,\]
where $\lambda$ is a weight parameter balancing the degree of the denoising and 
the fidelity to the input.
In terms of denoising quality,
TV denoising outperforms those simple methods (such as the median filtering)
% , which \tamal{I dont know `which' refers to what} may smooth away edges,
% during the denoising,
by preserving edges while reducing noise in flat areas. 

\paragraph{Wavelet thresholding.}
Wavelet thresholding filters work on the {\it frequency domain}.
Let images now be defined on the discrete 2D grid $I$
instead of the continuous $\OG$, and let $\Bcal=\Set{g_\aG}_{\aG\in A}$ be
an {\it orthonormal} basis for $\Real^{|I|}$.
The input image $v:I\to\Real$ can be decomposed into the form $v=\sum_{\aG\in A}v_\Bcal(\aG)g_\aG$,
where each coefficient $v_\Bcal(\aG)$ is the scalar product $\langle v,g_\aG\rangle$.
Then, each coefficient $v_\Bcal(\aG)$ is modified to $a(\aG)v_\Bcal(\aG)$,
where $a(\aG)$ is a value depending on $v_\Bcal(\aG)$.
Wavelet thresholding recovers a denoised image as 
$u=\sum_{\aG\in A}a(\aG)v_\Bcal(\aG)g_\aG$,
% with the modified coefficients. 
and different wavelet thresholding filters differ on how $a(\aG)$ is chosen
for each $v_\Bcal(\aG)$.
Specifically, {\it soft wavelet thresholding}~\cite{donoho1995noising}
sets $a(\aG)$ as:
\[
a(\aG)=\begin{cases}
% \begin{align*}\begin{split}
\;\frac{v_\Bcal(\aG)-\text{sgn}(v_\Bcal(\aG))\mu}{v_\Bcal(\aG)}, & |v_\Bcal(\aG)|>\mu,
% \text{ and }
\\
\; 0, & \text{otherwise,}
% \end{split}\end{align*}
\end{cases}\]
where $\mu$ is a threshold parameter.
{\it Translation invariant wavelet thresholding}~\cite{coifman1995translation}
further improves soft wavelet thresholding
by averaging the denoising results on all translations of the input image,
which could reduce the so-called {\it Gibbs effect}.

\paragraph{NL-means.}
The name NL-means~\cite{buades2005review} is short for `non-local' means
% , is proposed in
% with the naming 
which comes from its difference to those
`local mean' filters.
Given an input image $v$,
NL-means produces the following
denoised pixel $u(x)$ for a given position $x\in\OG$:
\[u(x)=\frac{1}{C(x)}\int_\OG e^{-\frac{L_a(x,y)}{h^2}}v(y)\mathrm{d}y,\]
in which $h$ is a filtering parameter, $C(x)$ is a normalizing factor,
and $L_a$ is a measure of 
dissimilarity of the Gaussian neighborhood of two pixels.
Specifically, let $G_a$ be a Gaussian kernel with standard deviation $a$,
we have 
\[L_a(x,y)=\int_{\Real^2}G_a(t)|v(x+t)-v(y+t)|^2\mathrm{d}t.\]
Hence, 
% the denoised pixel of NL-means 
$u(x)$ is the mean of 
the other pixels whose Gaussian neighborhood is similar to 
$x$.
% the given pixel.
% Because of its nature,
NL-means produces denoised images with greater clarity
and less detail loss compared to those local mean algorithms.

\paragraph{UINTA.}
UINTA~\cite{awate2006unsupervised} stands for `unsupervised, information-theoretic, adaptive filter'.
For each pixel $x$ and its neighborhood $N_x$, UINTA
uses stochastic gradient descent to 
reduce the entropy (expectation of negative log-probability) of the 
conditional probability density $P(X=x\given N_X=N_x)$,
where $P(X=x\given N_X=N_x)$ is estimated by
the Parzen-window non-parametric density estimation technique~\cite{hart2000pattern}
with an $n$-dimensional Gaussian kernel.
The design of UINTA is based on the observation that adding noise to the original signal 
% and 
% as two independent random variables
greatly increases the entropy. Hence, decreasing the entropy 
mitigates randomness in the observed PDF's and so noise can be reduced. 
Note that UINTA resembles NL-means in a sense that it also involves
comparing the neighborhood of pixels for computing a denoised value.

\newcommand{\maxgval}{M}
\section{{\sc PerSplat} algorithm}
\label{sec:persplat}
% \tao{for the header, I suggest we keep the bold face without sc. there is no combination of sc and bold face in Latex, and using an sc in the header looks really weird.}

We only describe the {\sc PerSplat} algorithm in 2D; its 3D version follows similar steps.
% Before presenting the details,
% we first briefly overview the algorithm.
There are two stages in
the algorithm,
% with the first stage traversing the grey-scale values of images 
% in {\it increasing} order (e.g., from 0 to 255)
% and the second stage traversing the values
% in {\it decreasing} order.
and since the second stage is a reversal of the first stage,
we only briefly overview the first stage before presenting the details.
In the first stage, we traverse the grey-scale values of the input image
in increasing order starting from zero.
For each value $s$ traversed, we consider pixels in the image
with grey-scale values no greater than $s$ (i.e., the {\it sublevel set})
and take the connected components of these pixels.
As $s$ increases, we track the following changes of connected components 
in the sublevel sets:
\begin{enumerate}
\item
A new connected component which does not correspond to any previous ones
is {\it created}; this happens at a {\it local minimum} in the image.
\item
Connected components grow larger.
% ({\sf iii}) 
\item
Several connected components {\it merge} into 
the same component.
% this happens at the {\it saddle} of the image.
\end{enumerate}

Consider a merging that happens, say at a grey-scale value $s$. 
We `splat' any connected component $\Csf$ that is getting merged
with size and {\it persistence length} (defined later)
no greater than some input parameters. By 
splatting a component, we mean to assign the grey-scale value $s$ to all its pixels.
Intuitively, the splatting 
suppresses those small `downward' (dark) bumps,
which serves as a noise reduction for the image.
Symmetrically, 
splatting in the second stage suppresses those small `upward' (bright) bumps;
see the discussion and figure presented later in the section
for more details.

\paragraph{Full details.}
% We now present the full details of the algorithm.
We formalize a 2D input image as a function $f:\Gsf\to[0,\maxgval]$,
where $[0,\maxgval]$ is an interval of integers with
$\maxgval$ usually equal to 255.
Moreover,
$\Gsf$ is the graph corresponding to the 2-dimensional grid of an image,
i.e.,
vertices of $\Gsf$ correspond to pixels in the image 
and connect to either 4 or 8 of its neighbors\footnote{In the experiments of this paper,
our implementation of the algorithm always connects a vertex 
to 4 neighbors for 2D and 6 neighbors for 3D.}.
We also have that
function values of $f$ on the vertices are 
grey-scale values of the corresponding pixels.

For any $s\in[0,\maxgval]$, define $f_s$ as the full subgraph of $\Gsf$
containing vertices whose function values are no greater than $s$.
We have that $f_s$ is a subgraph of $f_t$ whenever $s\leq t$.
Therefore, starting from $s=0$, $f_s$ keeps growing larger
% and larger
as $s$ increases and eventually equals the entire $\Gsf$.
For an $s$, we consider the {\it connected components} of $f_s$,
i.e., those maximal sets of vertices of $\Gsf$ in which
each pair admits a connecting path.
As we increase the value of $s$,
the following three types of events can happen, 
in which we pay attention to the first and the third one
(i.e., the {\it critical} events):
\begin{enumerate}
% ({\sf i}) 
\item
A new connected component in $f_s$ which has no correspondence in $f_{s-1}$
is {\it created}.
% ({\sf ii}) 
\item
Connected components of $f_{s-1}$ grow larger in $f_{s}$.
% ({\sf iii}) 
\item
Several connected components of $f_{s-1}$ {\it merge} into 
the same connected component in $f_{s}$. 
\end{enumerate}
% In our algorithm, we focus the creation and merging of components.

The algorithm requires two parameters $\specsize$ and $\perslen$ as inputs,
where $\specsize$ denotes the {\it maximum speckle size}
and $\perslen$ denotes the {\it maximum persistence length}
of a connected component that the algorithm can modify.
%The meaning of the two parameters becomes clear as we delineate further details. 
%Note that we have an emphasis on the impact of $\specsize$
%(often briefly called
%the {\it speckle size} parameter)
%to the algorithm
Note that we always set $\perslen=\infty$ in experiments so that the algorithm
can modify any connected component.

\begin{figure}
  \centering
  \includegraphics[width=\linewidth]{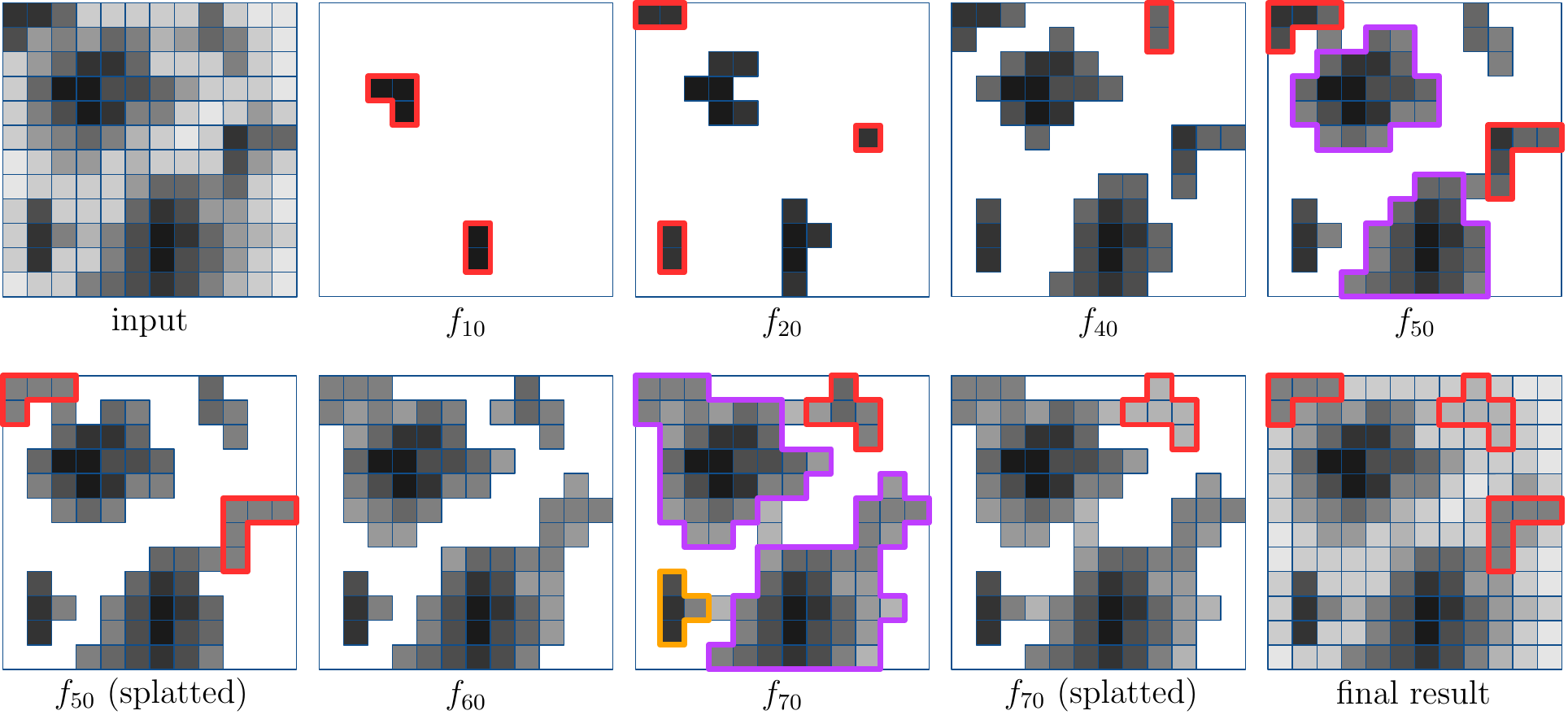}
   \caption{\textbf{An exemplar run of the first stage of {PerSplat}.}
   As $s$ increases, connected components can be created (e.g., red ones in $f_{10}$)
   or grow larger. 
   The connected components can also merge among which some get splatted (e.g., red ones in $f_{50}$)
   while some are preserved (e.g., purple and orange ones in $f_{70}$).}
    \label{fig:ppd2d}
\end{figure}

The algorithm contains two stages, where the first stage
increasingly enumerates $s$ from $0$ to $\maxgval$.
During the first stage, whenever a merging happens at $s$, 
for each connected component $\Csf$ of $f_{s-1}$ that merges with others,
we define the following:
\begin{itemize}
    \item Let $|\Csf|$ denote the {\it size} of $\Csf$, which is the number of vertices of $\Csf$.
    \item Let $\brm(\Csf)$ denote the {\it birth value} of $\Csf$,
which is the least function value of $\Csf$'s vertices.
    \item Let $\ell(\Csf)$ denote the {\it persistence length} of $\Csf$,
    which is defined as $\ell(\Csf)=|s-\brm(\Csf)|$.\footnote{
  Note that here we have an interval $[\brm(\Csf),s)$. However, this interval 
  is not exactly the same as an interval in a {\it persistence diagram}~\cite{EH10,edelsbrunner2000topological},
  which does not produce an interval for the merged connected component with the least birth value.}
\end{itemize}
If $|\Csf|$ is no greater than $\specsize$ 
and $\ell(\Csf)$ is no greater than $\perslen$,
we assign the value $s$ to all pixels corresponding
to the vertices of $\Csf$,
i.e., the connected component $\Csf$ is {\it splatted} to value $s$ 
(see Figure~\ref{fig:ppd2d}).
Note that the creation and merging of connected components actually construct
the {\it merge tree}
%\cite{morozov2013interleaving}
of $f$, which is a well-known tool in topological
data analysis (TDA).

The second stage is a reverse process of the first:
the value $s$ is enumerated decreasingly from $\maxgval$ to 0.
Since the input image has been modified in the first stage,
we let $g:\Gsf\to[0,\maxgval]$ denote the function corresponding to this modified image.
Furthermore,
we define $g^s$ as the full subgraph of $\Gsf$
containing  vertices whose function values (on $g$) are no less than $s$.
Therefore, as $s$ decreases,
connected components in $g^s$ can also get created, grow, or merge with others.
Note that birth value in this case is the greatest function value of the connected component.
% and the persistence length equals the birth value minus $s$.
We then modify the pixel values for a connected component whenever a merging happens,
which is similar to the previous stage.

As mentioned earlier, the first stage suppresses those small `downward' bumps for the image
and the second stage suppresses those small `upward' bumps.
The parameters $\specsize$ and $\perslen$ control how aggressive
the suppressing is,
in which trade-offs need to be made between noise filtering
and possible loss of details.
% We also note that 
% it is those bumps near the possible thresholds which contribute to
% a better thresholding result. 
Figure~\ref{fig:ppd2d} illustrates an exemplar run ($\specsize=5$, $\perslen=30$)
of the first stage,
where the input image is considered to have six dark spots.
Three smaller, less dark spots
are splatted (i.e., turn to grey; see the red squares in the final result),
leaving the remaining three spots which are more prominent preserved.
In the exemplar run, $f_{10}$ creates two connected components,
while $f_{20}$ and $f_{40}$ create three and one component(s) respectively;
the newly created components are marked by red squares.
Note that these connected components are constantly growing after being created.
In $f_{50}$, two pairs of connected components merge together in which the ones marked
by red squares are splatted because of their sizes and persistence lengths.
In $f_{70}$, all four connected components are merged where the one inside the red square
is splatted;
note that though the connected component inside the orange square has a size no greater than
$\specsize$, it is not splatted due to a longer persistence length than $\perslen$.

Finally, we observe the following property of {\sc PerSplat},
which confirms the consistency of choices made
to modify a connected component:

\vspace{\topsep}
% \begin{property*}
{\it\noindent
% \begin{itemize}
    % \item[] {\it 
For any node on the merge tree which is splatted by the algorithm, 
all its descendants are splatted. 
Equivalently, for any node which is not splatted, all its ancestors are not.
% } \end{itemize}
}
% \end{property*}
\vspace{\topsep}

\section{Results and discussion}\label{sec:results}
\renewcommand{\arraystretch}{1.25}

% \tao{We should decide how to present our results now with a new filtering (TV) coming in.
% I believe for the `varying mean distance' part, we can keep the current content, and
% for the `varying speckle size' part, we definitely should include TV.
% But I don't know whether we should include TV for the `overall result'.
% Also, if the tests on natural images are really good, I am
% wondering whether we really need the part `varying mean distance'...
% \textbf{We need to add the poisson noise for all the datasets in `overall result'
% and in `varying mean distance' maybe.}}\vspace{2em}

In this section, we 
% perform experiments to 
demonstrate
% the improvement on the binarization quality by 
the efficacy of our {\sc PerSplat} algorithm
by comparing it with two general-purpose filtering algorithms:
TV~\cite{rudin1992nonlinear,rudin1994total} and NL-means~\cite{buades2005review}.
% We chose TV and NL-means because of their popularity and
% also because they
% are known to produce reasonably good results~\cite{buades2005review}.
After applying the filters,
we use the global thresholding algorithm {\sc Otsu}~\cite{otsu1979threshold} 
as a benchmark for binarization. Note that although some other types of
binarization algorithms
(such as {\it locally adaptive thresholding}~\cite{iassonov2009segmentation}) may produce better
results in some cases, we choose {\sc Otsu} because of
its simplicity and {effectiveness}~\cite{iassonov2009segmentation}.
Since the focus of the paper is on image filtering for XCT data, 
using {\sc Otsu} is sufficient for our purposes\footnote{Note that
we independently apply {\sc Otsu} to {\it each slice} of the 3D data
in our experiments.}.
%We understand this can create directional bias. In many cases while performing {\sc Otsu} using stack histogram it was observed that few slices (or sub region) were completely white or black.  This is because a single threshold value is applied to all the images in the stack regardless of the average grey scale value of the slice. 
%{\sout{\tao{We tend to not make this stand out.}}.}}}

We briefly summarize our experiments as follows:
\begin{enumerate}
    \item 
    Using {\sc Otsu} for binarization,
    we compare 
    % the tortuosity computed from using {\sc Otsu} alone, and 
    the tortuosity computed from {\sc PerSplat}, TV, and NL-means filtering
    % and other filters (TV~\cite{rudin1992nonlinear,rudin1994total} and NL-means~\cite{buades2005review})
    to the {\it ground-truth} tortuosity for some synthetically 
    generated images; see Section~\ref{sec:syn} for details.
    
    \item 
    We apply {\sc PerSplat}, TV, and NL-means for filtering
    a natural 3D image 
    % which is then binarized by {\sc Otsu},
    and visually compare the binarization results by {\sc Otsu}; see Section~\ref{sec:nat} for details.
\end{enumerate}

% \tao{Before presenting the experiments, we first briefly describe our {\sc PerSplat}
% algorithm in Section~\ref{sec:persplat}.}
Note that
we apply our {\sc PerSplat} algorithm
% (detailed later)
on 3D grey-scale images in two ways in this paper: 

\begin{description}
    \item[{\rm{\sc PerSplat-P} (i.e., {\sc PerSplat-}{\it Pixelized}):}]
    % {\sc PerSplat-P} (i.e., {\sc PerSplat-}{\it Pixelized}):
    In this case, we process the 3D image {\it slice by slice} 
    % where the 3D input is treated as a stack of 2D pixelized slices. Each slice is processed 
    using the version of {\sc PerSplat} for 2D input. The outputs for all slices are then stacked to create a 3D filtered image.
    
    %\item 
    \item[{\rm{\sc PerSplat-V} (i.e., {\sc PerSplat-}{\it Voxelized}):}] 
    % {\sc PerSplat-V} (i.e., {\sc PerSplat-}{\it Voxelized}):
    In this case, {\it all slices together} are 
    % treated as a 3D voxelized image which is 
    processed by the version of {\sc PerSplat} for 3D input. 
    % The output is also a 3D filtered image.
\end{description}

\subsection{Results on synthetic datasets}
\label{sec:syn}

\begin{figure}
  \centering
  \includegraphics[width=1\linewidth]{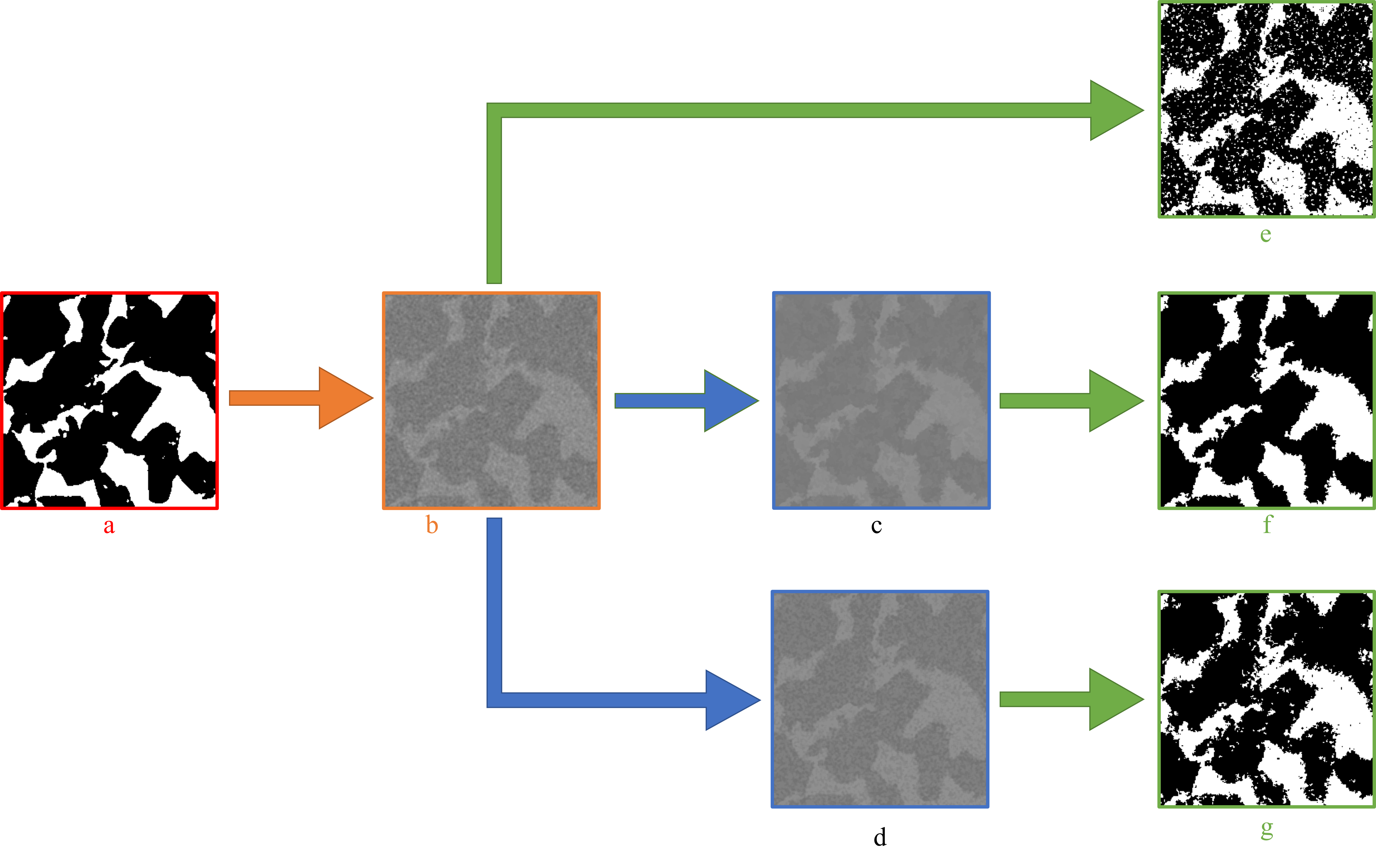} 
  \hspace{2em}
   \caption{%\sout{The blue color symbolises {\sc {\sc PerSplat}} algorithm and the green color symbolises binarization process.}\tao{It is best you denote this directly in the figure.}
   \textbf{Processing pipeline 
%   of the experiments 
   on synthetic datasets.}
   (a) A ground-truth slice of Synth1.
   (b)~Slice with synthetic noise added.
%   generated for the ground truth slice.
   (c) Noisy slice filtered by {\sc {\sc PerSplat}-P}.
   (d) Noisy slice filtered by {\sc {\sc PerSplat}-V}.
   (e,f,g) Binarized slices by {\sc Otsu}.}
   
   %\sout{(e) A 2D binarized data of (b) using otsu.
   %(f) A 2D binarized data of (c) using otsu.
   %(g) A 2D binarized data of (d) using otsu.}

    \label{fig:results2}
\end{figure}

% \paragraph{Synthetic dataset generation.}

The absence of ground truth for the XCT data \cite{Pietsch2018} makes it hard to 
evaluate the segmentation quality.
% compare the performance of the segmentation process. 
Hence, to address the issue of `missing' ground truth, we create synthetic images whose ground-truth values are known,
% \sout{Then, to justify the benefit of using {\sc {\sc PerSplat}},
% % for filtering,
% % in conjunction with {\sc Otsu}, 
% % a standard binarization scheme ({\sc Otsu}), 
% we compared the computed tortuosity values 
% by {\sc PerSplat} filtering
% and other filtering methods (TV and NL-means)
% with the known ground truth values.}
% As shown in Figure~\ref{fig:results2}, our experiments demonstrate improvements achieved by using {\sc {\sc PerSplat}}; 
% see Section~\ref{sec:results} for details of our experiments.
% One of the main difficulties we face when evaluating a
% segmentation approach for XCT images is the absence of ground truth.
% To sidestep this problem, 
% we synthetically generate datasets with known ground truth
% for our experiments 
% We generate the synthetic datasets
with the following process:
\begin{enumerate}
    \item Given a 3D XCT image, we first binarize the image with {\sc Otsu} and {\it take this binarized image as ground truth}, i.e., {we take  
     voxels with values greater than the threshold as in the solid phase
     and the rest as in the porous phase.}
   % and those with values less than are  in the porous phase.}
    
    %\sout{first establish a ground truth using  that we segment using {\sc Otsu} to define solid phase and porous phase.}
    \item We then introduce 
    % synthetic 
    noise to the established ground-truth image based on Gaussian distribution (see Section~\ref{sec:method} for further details).
    The introduction of Gaussian noise
    is justified by the fact that
    nearly all images we obtain through XCT have histograms
    akin to a mixture of Gaussians as shown in Figure~\ref{fig:results1}.
    % \tao{briefly explain the two parameters (mean dis, var) in the syn data generation
    % to prepare people for the later results.}}
    
    %\sout{All images that we obtain through XCT measurements typically have some kind of Gaussian distribution of grey-scale values that may overlap, as show in Figure~\ref{fig:results1}. }
    
    \item We also apply noise to the image based on Poisson distribution.
    The rationale {behind} the Poisson noise is as follows. 
    Typically, noise in XCT images is caused by various factors such as the number of scans performed, the object being measured, and the data processing {methodology} employed. One such factor is called `shot noise' or `quantum noise', which is introduced because photons in the X-ray beam 
    % follow the `quantum distribution'. These photons 
    follow a spatial distribution according to the Poisson law \cite{XCTnoise}.
    
    \item
    % After the noise is applied, 
    Finally, for each voxel, we apply a nearest neighbor averaging in three dimensions to account for {averaging of} the XCT signal that happens at the interfaces.

\end{enumerate}

% \paragraph{Experimental pipeline.}
% \label{sec:pipeline}

%\begin{itemize}
   % \item 

%\sout{In other words, the topological analysis is done simultaneously in all the directions.}
%\end{itemize}

% We set up the following pipeline for the experiments as shown in Figure~\ref{fig:results2}:

% \begin{enumerate}
%     \item First, we apply our {\sc PerSplat} algorithm (detailed later) for filtering the test images.
%     \item Then, we binarize the filtered images using {\sc Otsu} and calculate tortuosity
%     on the binarized images using a MATLAB open-source application {\sc TauFactor} \cite{Cooper2016}. The tortuosity is then compared to the ground truth values. 
% \end{enumerate}

% To show the improvements on binarization by using {\sc PerSplat},
% we adopt the following methods for binarizing our {3D input} data (see Figure~\ref{fig:results2}):

% \begin{description}
%     \item[{\rm{\sc Otsu}:}] 
%     This is the approach which directly uses {\sc Otsu} for binarization.\anand{For our experiments we have used {\sc Otsu} per slice.}
%     \item[{\rm{\sc PerSplat-P}+{\sc Otsu}:}]
%     We first apply {\sc PerSplat-P} for filtering the input and
%     then use {\sc Otsu} for binarization.
%     \item[{\rm{\sc PerSplat-V}+{\sc Otsu}:}] This is the same as the previous method 
%     with the only difference that {\sc PerSplat-V} is now used for the filtering.
% \end{description}

In Figure~\ref{fig:results2}, we show 
the processing pipeline 
on the synthetic datasets with {\sc PerSplat-P}
and {\sc PerSplat-V} for filtering.
After the binarization,
we calculate the tortuosity for the binarized images
using a MATLAB open-source application {\sc TauFactor} \cite{Cooper2016}. 
% The tortuosity is then compared to the ground truth values. 
 
%and compare the tortuosity determined by the \emph{tortuosity} computed from the ground truth and the cleaned image. 
%We binarize both the ground truth image and the cleaned image by the well known {\sc Otsu} algorithm before computing the tortuosity. 

\paragraph{Overall results.}
% In our experiments, 
Based on different ground truths,
we generate synthetic datasets with different degrees of overlaps in the histograms.
%{\Anand {For me to do: Reminding about histogram overlap and showing we created intentionally images with various overlap to see how the {\sc PerSplat} works on them}
To control the histogram overlap,
we adjust the mean and variance of the Gaussian noise for the two phases:
when the mean for the two phases
is close or the variance is large,
overlap in the histogram is significant (see Section~\ref{sec:method} for more details). 
% \tamal{This sentence is odd. When we say `closer' `larger', compared to what? Maybe, we should rewrite it without the comparatives.}
% (see Figure~\ref{fig:peakdis} 
% which shows histograms of synthetic datasets 
% with various mean distances
% for the same ground truth).

Table~\ref{tab:data-setting} provides details of the generated synthetic datasets,
where we list the data sources 
(i.e., the data samples from which we derived our ground truth),
black phase fractions of the ground-truth images,
and the Gaussian noise parameters for the two phases. 
The ground truths for Synth1 and Synth2 
are extracted from (different regions of) the XCT data of porous templated polymer.
%\tao{In the table, you use `Polymer', make it consistent.} \anand{keep it porous polymer} made in our lab\tao{`our' is not the most correct word. maybe `lab where some authors of this paper belong'?} \anand{I made the sample.}. 
The ground truths for Synth3 and Synth4 
are extracted from the tomographic data of a commercial LIB anode \cite{vwoodseth, Muller2018}.
% \tamal{I think it is a good idea to use `ground-truth' because you are using it as an adjective.}
The two sets of samples, porous polymer and LIB anode, are selected because of their difference in the phase fractions.
Note that the histogram of Synth2 has more overlap than that of Synth1
as exhibited by Table~\ref{tab:data-setting};
similar difference holds for Synth3 and Synth4.

%the mean of the white phase “active material” and the mean of the black phase “porous region” are about 15 units apart from each other with the same standard deviation.
%Synth2 was extracted from the same tomographic data of porous templated polymer but the mean difference between the black and the white phase is 7 units with standard deviation being 35. 
%Synth3 was extracted from the tomographic data of commercial LIB anode [6] and the mean difference and standard deviation is similar to Synth1.

% Similarly, Synth4 was extracted from the same tomographic data as Synth3 but with the Gaussian values as used in Synth2. 
%\tao{Now we have already had the data configuration table with all the detail; I also added the data source (please check if this is the right name).So this paragraph should remove all the detailing which is now unnecessary, but introduce some general info. Also, how the data are generated are not mentioned. maybe you want to put to the method section, but here at least you should have a brief intro so that people can keep reading.}

\begin{table}[!thbp]
\centering
\caption{\textbf{Statistical setting for generated synthetic data.} 
The ground truths for Synth1 and Synth2 are extracted from porous polymer. 
The ground truths for Synth3 and Synth4 are extracted from LIB anode. 
Active material is assigned as white phase while porous region is assigned as black phase.} 
%exemplar samples\tao{What does `exemplar sample' mean?}
\label{tab:data-setting}
% \resizebox{\textwidth}{!}{
% \begin{tabular}{|c|c|c|c|c|c|c|}
\begin{tabular}{ccccccc}
% \hline
\toprule
  \multirow{2}{*}{\makecell{\bf Dataset}}
  & \multirow{2}{*}{\bf Data source} &
  \multirow{2}{*}{\makecell{\bf Black phase\\\bf fraction }} &
  \multicolumn{2}{c}{\bf White phase} & \multicolumn{2}{c}{\bf Black phase} \\
  \cmidrule{4-7}
  & & & \makecell{\bf Mean } &
    % \begin{tabular}[c]{@{}c@{}}Mean\\White Phase\end{tabular} & 
    \makecell{\bf Var} & 
    % \begin{tabular}[c]{@{}c@{}}Std\\White Phase\end{tabular} & 
    \makecell{\bf Mean} & 
    % \begin{tabular}[c]{@{}c@{}}Mean\\Black Phase\end{tabular} & 
    \makecell{\bf Var} 
    % \begin{tabular}[c]{@{}c@{}}Std\\Black Phase\end{tabular} 
\\ \midrule
 Synth1 & porous polymer & 66\%  & 100 & 25 & 85 & 25 
\\ 
% \hline
 Synth2 & porous polymer & 75\% & 87 & 35 & 80 & 35 
\\ 
% \hline
 Synth3 & LIB anode  & 35\% & 100 & 25 & 85 & 25 
\\
% \hline
 Synth4 & LIB anode & 35\% & 87 & 35 & 80 & 35 
\\ \bottomrule
\end{tabular}%
% }
\end{table}

\begin{table}[]
\centering
\caption{\textbf{Microstructure characteristics and binarization accuracy {computed}.}
% \textbf{Tortuosity and binarization accuracy calculated.} 
% For each filter, 
% tortuosity of all directions, characteristic tortuosity
% (the harmonic mean of tortuosity in the three directions),
% and pixel accuracy of the binarized images are listed. 
% For each dataset,
For each type of microstructure characteristics,
we list 
% the ground truth-value and
the difference to the ground-truth value
% (tortuosity of all directions and the characteristic tortuosity)
calculated from each filter.
% We also list the pixel accuracy of the binarization for each filter.
% Tortuosity values closest to the ground truth
% and highest binarization accuracy 
The best values are underlined.}
% Note that characteristic tau factor is the harmonic mean of tau factor in three direction.
\label{tab:finalresults}
{\small\begin{tabular}{crrrrrrrr}
\toprule
\multirow{2}{*}{\bf Data} & \multicolumn{1}{c}{\multirow{2}{*}{\bf Methods}} & \multicolumn{3}{c}{\bf Tortuosity}             & \multicolumn{1}{c}{\multirow{2}{*}{\makecell{\bf Charac.\\\bf Tor.}}} & \multicolumn{1}{c}{\multirow{2}{*}{\makecell{\bf Pixel\\\bf Accur.}}} & \multicolumn{1}{c}{\multirow{2}{*}{\makecell{\bf White\\\bf Vol.}}} &	\multicolumn{1}{c}{\multirow{2}{*}{\makecell{\bf White\\\bf Surf. Area}}} \\
	
\cmidrule{3-5}
                             & \multicolumn{1}{c}{}                         & \makecell{1}            & \makecell{2}            & \makecell{3}            & \multicolumn{1}{c}{}                                                    & \multicolumn{1}{c}{}                                            \\
\midrule
\multirow{7}{*}{Synth1}     & Ground Truth                                 & 1.668       & 1.668        & 1.890        & 1.736                                                                   & -                    & 6058513 &	1.03E+06                                           \\
\cmidrule{2-9}
                             & No Filtering                                  & +0.272       & +0.274       & +0.274       & +0.274                                                                  & 90.45\%                            & +586589 &	+2.32E+06                 \\
                             & {\sc PerSplat-P}                                  & \underline{+0.037} & \underline{+0.036} & \underline{+0.138} & \underline{+0.065}                                                            & \underline{95.95\%}                              & +99754	&\underline{+2.98E+05}                      \\
                             & {\sc PerSplat-V}                                  & +0.146       & +0.146       & +0.192       & +0.160                                                                  & 93.84\%           &+379728	&+1.12E+06                                                \\
                             & NL-means-2D                                  & +0.272       & +0.274       & +0.274       & +0.274                                                                  & 90.45\%             &+586492	&+2.32E+06                                              \\
                             & NL-means-3D                                  & +0.271       & +0.272       & +0.273       & +0.273                                                                  & 90.48\%                 &+579665	&+2.31E+06                                          \\
                             & TV                                         & $-$0.127       & $-$0.130       & $-$0.184       & $-$0.144                                                                  & 93.70\%             &\underline{+79869}	&$-$3.03E+05                                              \\
\midrule
\multirow{7}{*}{Synth2}     & Ground Truth                                 & 1.392        & 1.391        & 1.345        & 1.376                                                                   & -                                                       &4213761	&7.58E+05        \\
\cmidrule{2-9}
                             & No Filtering                                  & +1.239       & +1.233       & +1.203       & $-$1.225                                                                  & 64.02\%                          &+3878836	&+6.96E+06                                 \\
                             & {\sc PerSplat-P}                                  & +0.635       & +0.646       & +2.027       & $-$0.966                                                                  & 72.74\%                &+3600816	&+4.87E+06                                           \\
                             & {\sc PerSplat-V}                                  & +1.427       & +1.425       & +1.665       & $-$1.503                                                                  & 62.29\%                       &+4527724	&+6.74E+06                                    \\
                             & NL-means-2D                                  & +1.239       & +1.233       & +1.203       & $-$1.225                                                                  & 64.02\%                            &+3878766	&+6.96E+06                               \\
                             & NL-means-3D                                  & +1.202       & +1.198       & +1.172       & $-$1.191                                                                  & 64.26\%                              &+3811400	&+6.96E+06                             \\
                             & TV                                         & \underline{$-$0.064} & \underline{$-$0.058} & \underline{$-$0.033} & \underline{+0.051}                                                            & \underline{93.85\%}                                            &\underline{+303944}	&\underline{$-$4.93E+04}         \\
\midrule
\multirow{7}{*}{Synth3}     & Ground Truth                                 & 3.078        & 2.701        & 3.496        & 3.058                                                                   &     -                                                   &5530602	&1.73E+06         \\
\cmidrule{2-9}
                             & No Filtering                                  & +0.171       & +0.244       & $-$0.140       & $-$0.116                                                                  & 82.75\%                                 &$-$568356	&+9.78E+05                          \\
                             & {\sc PerSplat-P}                                  & +0.189       & \underline{+0.063} & +0.570       & $-$0.225                                                                  & \underline{86.22\%}                                       &\underline{+76852}	&\underline{+2.26E+04}              \\
                             & {\sc PerSplat-V}                                  & \underline{+0.151} & +0.202       & \underline{$-$0.078} & $-$0.111                                                                  & 84.01\%                                                   &$-$399765	&+7.32E+05        \\
                             & NL-means-2D                                  & +0.171       & +0.245       & $-$0.140       & $-$0.116                                                                  & 82.76\%         &$-$568284	&+9.78E+05                                                  \\
                             & NL-means-3D                                  & \underline{+0.151} & +0.232       & $-$0.160       & \underline{$-$0.099}                                                            & 82.65\%                                                      &$-$583979	&+9.90E+05     \\
                             & TV                                         & $-$0.803       & $-$0.328       & $-$1.443       & +0.832                                                                  & 74.63\%        & +90793	&$-$1.12E+06                                                  \\
\midrule
\multirow{7}{*}{Synth4}     & Ground Truth                                 & 3.078        & 2.701        & 3.496        & 3.058                                                                   & -                                              &5530602	&1.73E+06                 \\
\cmidrule{2-9}
                             & No Filtering                                  & $-$0.402       & $-$0.080       & $-$0.821       & +0.400                                                                  & 62.90\%             &$-$1216905	&+2.33E+06                                              \\
                             & {\sc PerSplat-P}                                  & $-$0.365       & $-$0.112       & +2.609       & $-$0.208                                                                  & 69.39\%                                                         &$-$651080	&+1.48E+06  \\
                             & {\sc PerSplat-V}                                  & \underline{$-$0.296} & \underline{+0.026} & \underline{$-$0.335} & \underline{+0.180}                                                            & 63.88\%                                               &$-$962743	&+2.23E+06         \\
                             & NL-means-2D                                  & $-$0.403       & $-$0.079       & $-$0.820       & +0.400                                                                  & 62.91\%                    &$-$1216840	&+2.33E+06                                       \\
                             & NL-means-3D                                  & $-$0.405       & $-$0.077       & $-$0.819       & +0.400                                                                  & 62.90\%                &$-$1213542	&+2.34E+06                                           \\
                             & TV                                         & $-$0.816       & $-$0.555       & $-$1.102       & +0.795                                                                  & \underline{71.82\%}   &\underline{$-$439236}	&\underline{$-$8.00E+05}  \\                   
\bottomrule
\end{tabular}}
\end{table}

In Table~\ref{tab:finalresults},
we show the following values computed with the different filters
for each dataset in Table~\ref{tab:data-setting}:
\begin{itemize}
    \item 
tortuosity in the three directions and
the characteristic tortuosity (the harmonic mean of tortuosity in the three directions);
    \item
volume and surface area of the white phase, which are {a subset of the Minkowski functionals~\cite{legland2007computation}};
    \item
binarization accuracy, which is the percentage of voxels correctly classified.
\end{itemize}

% calculated for the datasets listed in Table~\ref{tab:data-setting}.
% We applied the different filters mentioned
% for binarization and 
We also underline the best values achieved in Table~\ref{tab:finalresults}.
% Note that `characteristic tortuosity' is the harmonic mean of tortuosity in the three directions.
Note that `NL-means-2D' means that 
NL-means is applied independently to each slice
and `NL-means-3D' means that NL-means
is applied to the entire stack at once.
We fix the parameters of the filtering algorithms
for all the datasets in Table~\ref{tab:finalresults}:
\begin{itemize}
    \item for {\sc PerSplat-P}, we use a speckle size $\specsize=54$;
    \item for {\sc PerSplat-V}, we use a speckle size $\specsize=500$;
    \item for TV, we use a regularization {\it weight} of 0.5;
    \item for NL-means-2D and -3D, we use the default parameters 
    of the implementation provided by {\tt scikit-image}~\cite{van2014scikit}.
\end{itemize}

From Table~\ref{tab:finalresults},
we observe that {\sc PerSplat-P} or {\sc PerSplat-V} 
are 
consistently providing the most accurate tortuosity values.
The only exception is Synth2, whose best values
are provided by TV. However, on Synth2, the tortuosity values provided
by {\sc PerSplat} are still comparable to those provided by NL-means.
% Note that on Synth3, although NL-means-3D provides the best characteristic
% tortuosity, we observe that this is due to NL-means-3D made more {\it over}-estimation
% in direction 2 and more {\it under}-estimation in direction 3 compared to {\sc PerSplat-V},
% so that the errors are cancelled in the harmonic mean.
% Hence, the filter on Synth3 is {\sc PerSplat-V}. 
As for the volume and surface area of the white phase shown in Table~\ref{tab:finalresults},
we observe that
TV provides the best results with {\sc PerSplat-P} 
performing comparably well, {whereas} \mbox{\sc PerSplat-P} 
and \mbox{{\sc PerSplat-V}} consistently outperform
NL-means. 
% The reason for TV being the best {\red on this type of microstructure characteristics}
% is that TV tends to produce \sout{more smooth} images with less
% speckles. {\red This favors the computation of the volume and surface area of the white phase.} \anand{reword this}\tamal{Yes, I dont know what it means.}
% However,
% the smoothing by 
% TV may also introduce
% some structural errors in the binarized images.
% This is evident from the tortuosity {value depicted} in Table~\ref{tab:finalresults}
% as well as the visual presentation in Figure~\ref{fig:synth1},
% which is discussed later.

% among all the methods.
% \sout{From Table~\ref{tab:finalresults},
% we observe that the best
% values for nearly all datasets were achieved by using {\sc PerSplat-P}
% except Synth4 whose best tortuosity values were achieved by using {\sc PerSplat-V}.
% Therefore, results in Table~\ref{tab:finalresults} 
% are consistent with our claim that {\sc PerSplat} improves
% the thresholding quality of approaches like {\sc Otsu}.
% Note that we fix speckle size $\specsize$ for all datasets
% when running {\sc PerSplat-P} or {\sc PerSplat-V} for comparison purposes;
% %and do not try to cherrypick $\specsize$
% see experiments presented later in the section
% for influence of the $\specsize$ parameter
% on binarization.}

\begin{figure}[!tbhp]
  \centering
  \includegraphics[width=1\linewidth]{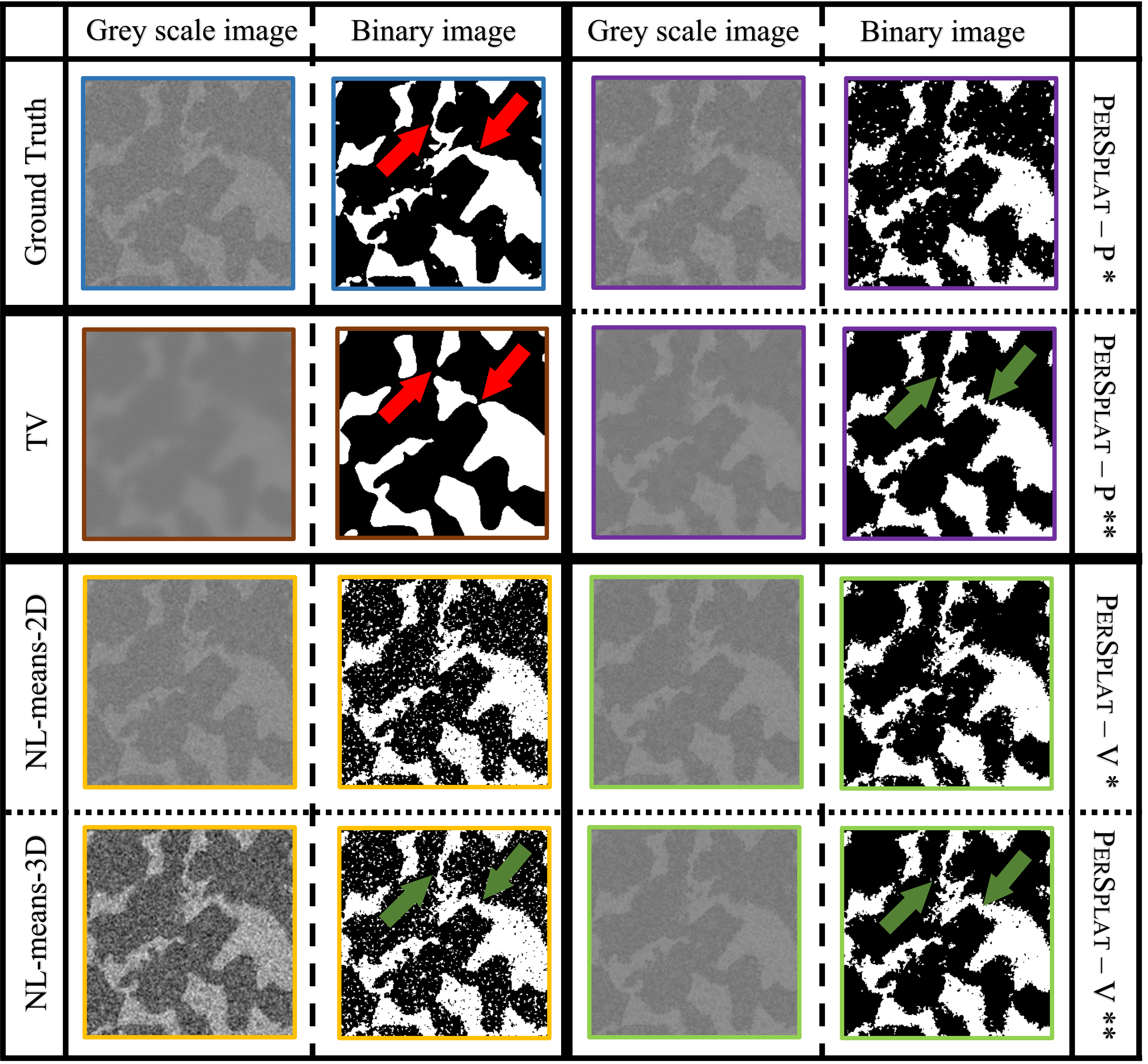} 
  \hspace{2em}
   \caption{\textbf{Grey-scale images filtered by all methods and their binarized version
   for one slice of Synth1, with the unfiltered slice
   and the ground-truth binarization also shown.} 
   Slices processed by different filters are marked with different colors.
%   Blue border marks ground truth data. 
%   Brown border marks data filtered by TV.
%   Orange border marks data filtered by NL-means. 
%   Purple border marks data filtered by 
   {\sc PerSplat-P$^*$} has $\specsize=10$; {\sc PerSplat-P$^{**}$} has $\specsize=54$;
%   Green border marks data filtered by 
   {\sc PerSplat-V$^*$} has $\specsize=500$; {\sc PerSplat-V$^{**}$} has $\specsize=14659$.}
%   Red arrows and green arrows highlight the area of interest.}
   \label{fig:synth1}
\end{figure}

% \sout{Figure~\ref{fig:results2} illustrates the improvements on segmentation quality
% introduced by using {\sc PerSplat}, where Figure~\ref{fig:results2}a 
% shows a 2D slice of the ground truth of Synth1
% and Figure~\ref{fig:results2}b shows the corresponding
% slice of Synth1 with synthetic noise.
% Using {\sc Otsu} alone for segmentation resulted in a speckly image due to overlaps in the histogram (Figure~\ref{fig:results2}e). 
% % Then we ran {\sc PerSplat-P} algorithm which resulted in a new grey-scale image (Figure~\ref{fig:results2}c). 
% In contrast,
% after using {\sc PerSplat-P} and {\sc PerSplat-V} for filtering
% (Figure~\ref{fig:results2}c and \ref{fig:results2}d),
% % and then applying {\sc Otsu}, 
% the resulting images segmented by {\sc Otsu} (Figure~\ref{fig:results2}f and \ref{fig:results2}g) 
% preserved the important regions with less speckle
% (see also Figure~\ref{fig:peakdis}).}
For one slice of Synth1,
we show in Figure~\ref{fig:synth1}
the corresponding grey-scale images
filtered by methods in Table~\ref{tab:finalresults}
and the binarized version by {\sc Otsu};
the original unfiltered slice of Synth1
and the ground-truth binarization
are also shown.
We also include
in Figure~\ref{fig:synth1}
the results from {\sc PerSplat}
using two speckle sizes ($10\text{ and }14659$)
different from the ones (54 and 500) used to generate Table~\ref{tab:finalresults}.
From the binarized image with TV filtering in Figure~\ref{fig:synth1}, 
% tends to be excessively smoothed so that
% , which causes some
% critical structural inaccuracy in the binarized slice. For example, 
we observe that
the white connected component marked by the red arrows 
% in the ground truth binarization 
is inaccurately disconnected.
% in
% the binarized image from TV filtering.
To the contrary, {\sc PerSplat} and NL-means both preserve the connected component as indicated by the green arrows in Figure~\ref{fig:synth1}.
Also note that
the binarized images with NL-means filtering in Figure~\ref{fig:synth1}
have many
speckles which the ground truth does not have.
In contrast, the speckles are 
largely avoided by {\sc PerSplat} (especially {\sc PerSplat-P} with speckle size 54).

\paragraph{Varying speckle size.}

\begin{figure}[!tbhp]
  \centering
  \subfloat[]{\includegraphics[width=.5\linewidth]{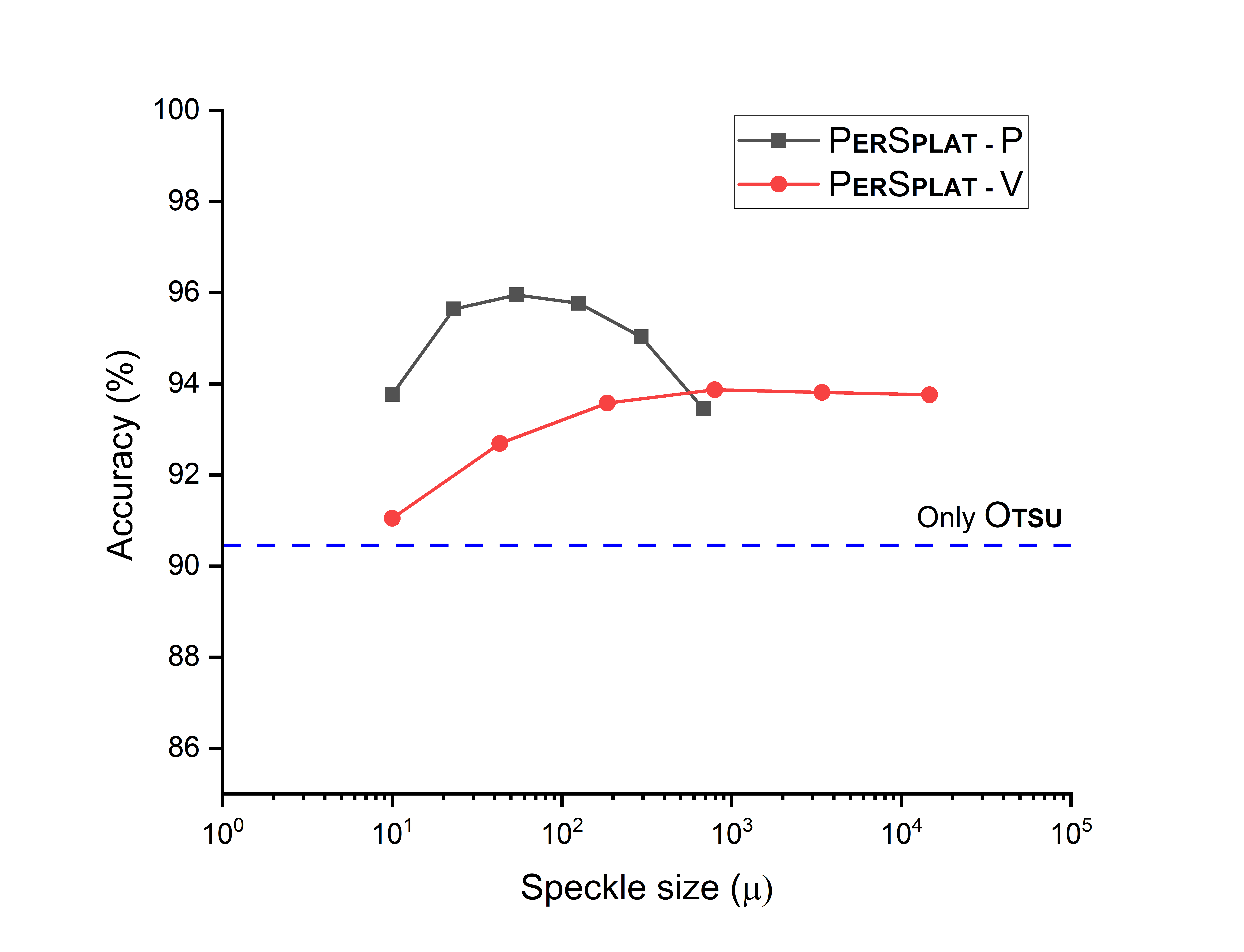}}
  \subfloat[]{\includegraphics[width=.5\linewidth]{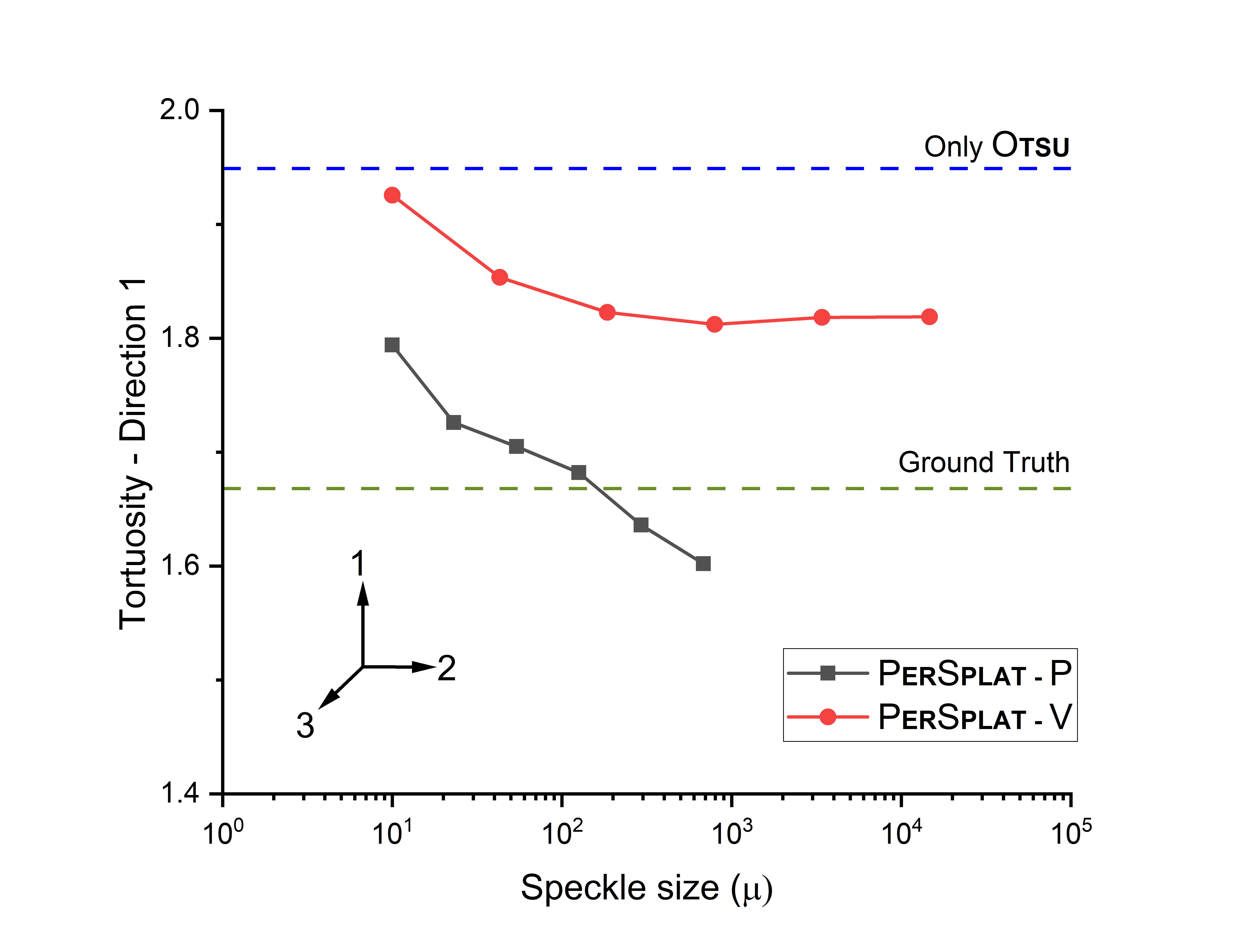}}\\
  \subfloat[]{\includegraphics[width=.5\textwidth]{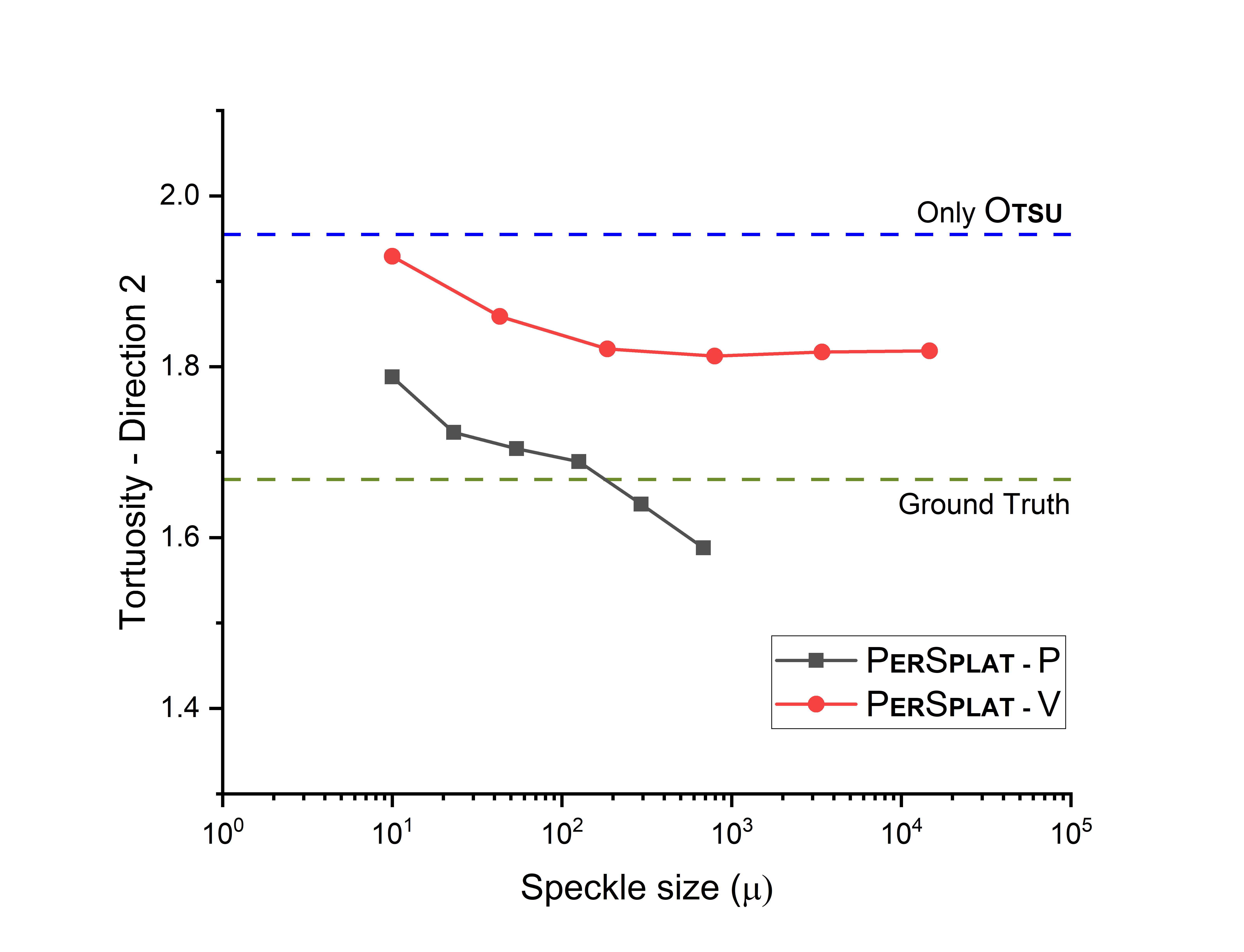}}
  \subfloat[]{\includegraphics[width=.5\textwidth]{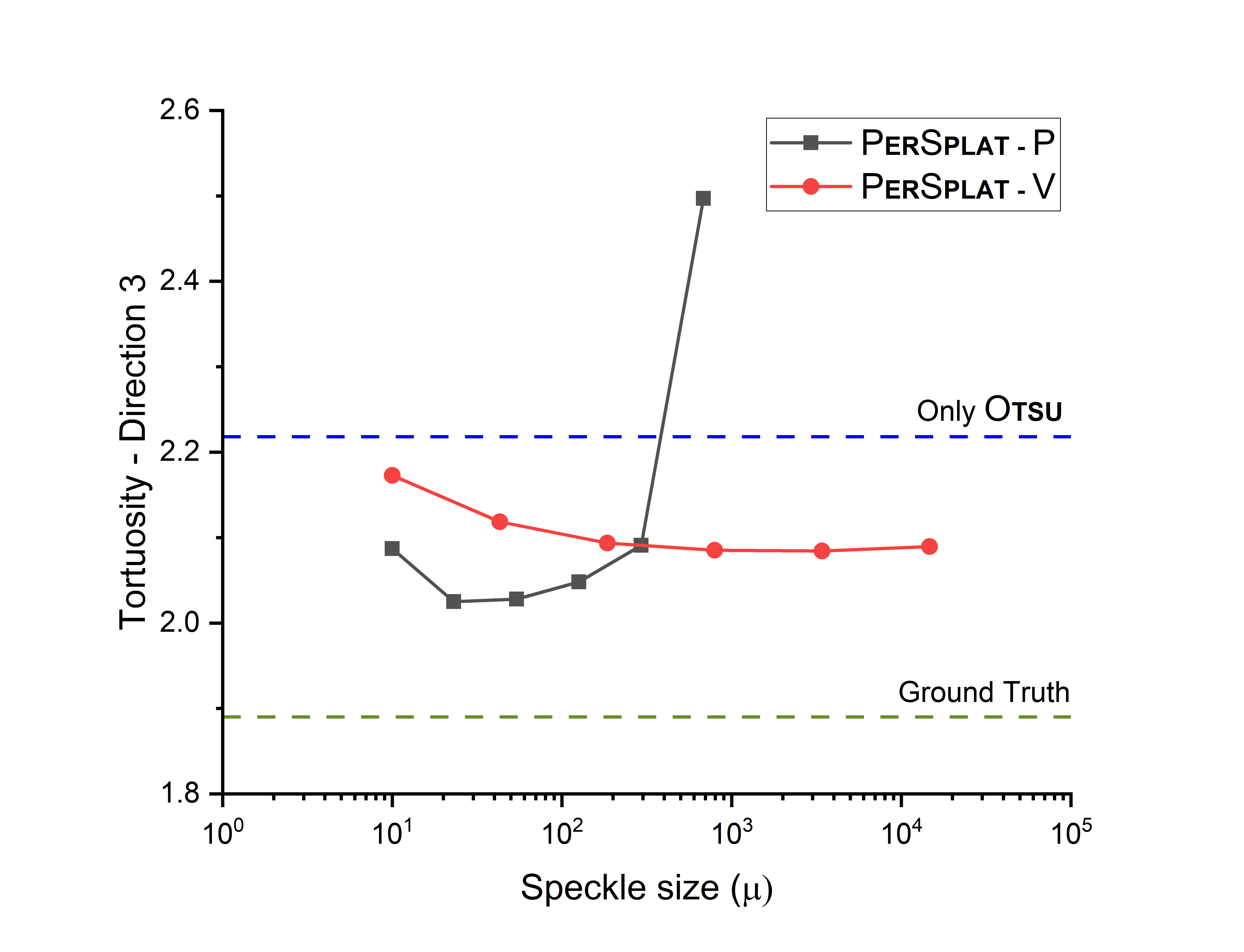}}
  \caption{\textbf{Microstructure analysis with various speckle size ($\specsize$) parameters for {\sc PerSplat}}.
  The input dataset is Synth1 and
  horizontal axes for the $\specsize$ parameter
  are re-scaled by logarithm function.
  (a)~Binarization accuracy as a function of $\specsize$. 
  (b-d) Tortuosity as a function of $\specsize$ in the three directions.}
  \label{plot1}
\end{figure}

Recall that,
with {\sc PerSplat}, the user needs to choose a speckle size parameter $\specsize$
as input.
To provide guidance on the parameter choosing,
% while treating the data in 2D or 3D. 
in Figure~\ref{plot1}, 
% plot b, c and d, 
we show how various $\specsize$ parameters 
impact the binarization accuracy 
and the computed tortuosity. The input dataset for Figure~\ref{plot1} is Synth1
and the speckle sizes 
exponentially range from 10 to 682 for {\sc PerSplat-P},
from 10 to 14659 for {\sc PerSplat-V}.
% in Figure~\ref{plot1} to 
% be about {\red 30\% of the pixel/voxel count of the input}.
% since slices are processed independently in {\sc PerSplat-P},
% the maximum $\specsize$ for {\sc PerSplat-P} is smaller
% than that for {\sc PerSplat-V}.
% (in all three directions).
% of the segmentation results. 
From Figure~\ref{plot1}, 
we observe that
the change of speckle size has a more significant effect on tortuosity when using {\sc {\sc PerSplat}-P}, 
while tortuosity appears to be more stable at various speckle sizes when using {\sc {\sc PerSplat}-V}. 
This is a result of the additional dimension of connectivity introduced in 3D.
% \footnote{
To understand the phenomenon, imagine running {\sc PerSplat} on 2D and 3D images:
as the value $s$ increases/decreases (see the description of the algorithm),
the sizes of the connected components grow much faster in 3D than in 2D
because of the additional dimension of connectivity.
In 3D, the connected components quickly grow very large; to splat them,
we need a large $\specsize$ parameter.
% Thus, any small $\specsize$ parameter has little effects on the 
% results.
This makes nearly the same components to be splatted
for a wide range of $\specsize$
when running {\sc PerSplat} on 3D images ({\sc PerSplat-V}).
% }. 
%\sout{In other words, when running {\sc {\sc PerSplat}-P}, the surfaces of an object in direction 3 would get splatted because of speckle size criteria. On the other hand, when running {\sc {\sc PerSplat}-V} the same object will be preserved at various speckle size. This makes {\sc {\sc PerSplat}-V} more stable and reliable than {\sc {\sc PerSplat}-P} in case of a 3D image.}  
% From Figure~\ref{plot1}, we also observe that with the right $\specsize$ (around 100), 
% {\sc PerSplat-P} achieved the best tortuosity values
% in all directions;
% this consistently happened in most of our experiments
% on synthetic datasets
% (some not shown in the paper).

\begin{figure}[!tbhp]
  \centering
% \begin{subfigure}
    \subfloat[]{\includegraphics[width=0.4\linewidth]{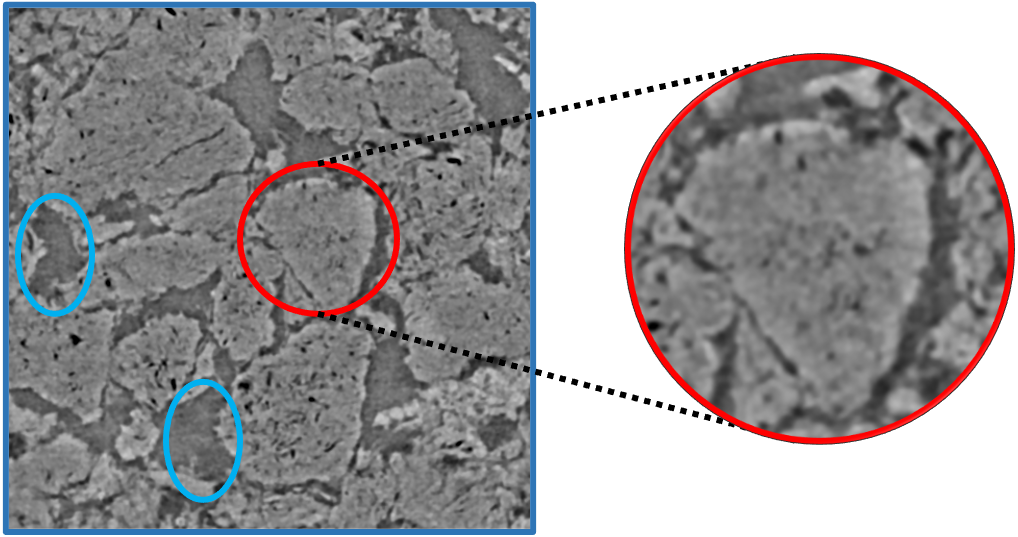}}
    % \label{fig:nat1}
% \end{subfigure}

% \begin{subfigure}
    \subfloat[]{\includegraphics[width=1\linewidth]{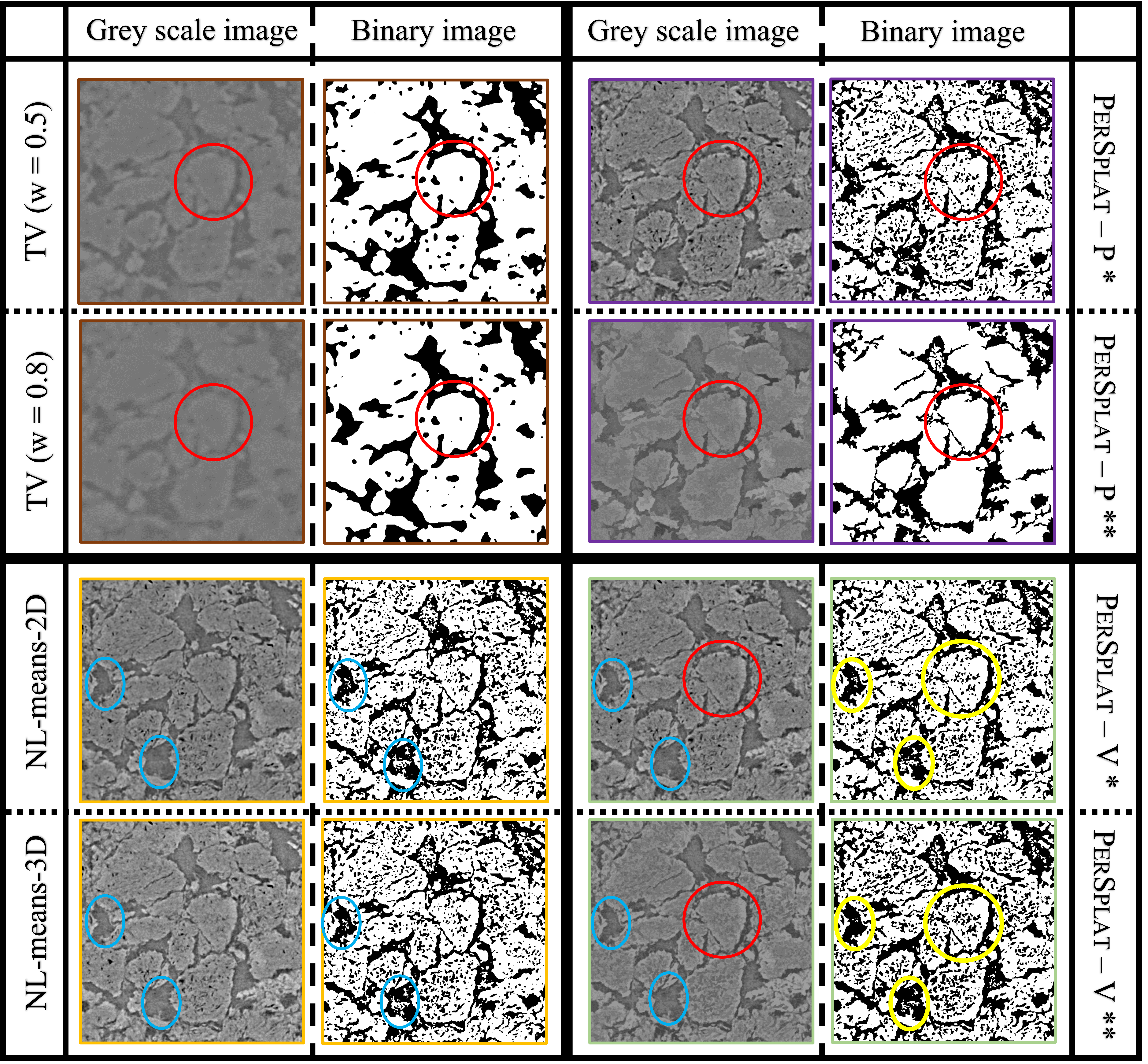}}
    \caption{
    % \textbf{A slice 
    % % \sout{(50th)} 
    % from a natrual 3D raw image gathered from an XCT experiment, their respective filtered slices, and binary slices.} 
    \textbf{One slice of a natural XCT dataset processed by different filters.}
    (a) The input unfiltered slice.
    % {\red and its binarized result by {\sc Otsu}}. 
    (b) Filtered and binarized slices, 
    with the slices processed by different filters marked with different colors.
    % Brown border marks data filtered by TV. 
    % Orange bordermarks data filtered by NL-means.  
    % Purple border marks data filtered by PerSplat-P$^*$ (with $\specsize= 10$) and PerSplat-P$^{**}$ (with $\specsize= 293$).  Green border marks data filtered by PerSplat-V$^*$ (with $\specsize= 10$)
    % and PerSplat-V$^{**}$ (with $\specsize= 14659$). 
    {\sc PerSplat-P$^*$} has $\specsize$ = $10$; {\sc PerSplat-P$^{**}$} has $\specsize$ = $293$;
    {\sc PerSplat-V$^*$} has $\specsize$ = $10$; 
    {\sc PerSplat-V$^{**}$} has $\specsize$ = $14659$.}
    % \tao{You should also have a binarized image for the unfiltered slice. Anand: I dont think that is necessary} \tao{It would show the improvement.}} 
    %Red and yellow circles are area of interest.}
    \label{fig:nat2}
% \end{subfigure}

\end{figure}

\subsection{Results on natural datasets}
\label{sec:nat}

In Figure~\ref{fig:nat2}, we show the filtered images by
the three methods (TV, NL-means, {\sc PerSplat})
and their {\sc Otsu}-binarized results
for one slice of a natural XCT dataset. 
The natural XCT dataset is gathered from a tomography experiment from an open source~\cite{Pietsch2017}.
%\sout{We took the following steps for the data analysis: (a) gathered raw tomography data from an open source, a 50th slice from the data is shown in Figure~\ref{fig:nat2}, (b) this data (entire 3D data) was then filtered using three different algorithms separately, and (c) finally, was binarized using {\sc{Otsu}}.}
%\tao{Most of the information of the above sentence is implicit and don't need to be mentioned. Maybe merge `gathered raw tomography data from an open source' to the sentence before?}
By visually examining the results in Figure~\ref{fig:nat2}, 
we notice that TV produces overly smoothed images in which 
many important details (e.g., ones inside the white phase marked by the red circle) are lost. 
%\sout{It also seems to interrupt the connectedness of the phases as it was shown previously with synthetic data (synth1).} 
In contrast, NL-means (2D and 3D) manage to preserve 
most of the important details inside the white phase marked by the red circle.
However, with NL-means filtering, 
some other important regions 
% from the black phase
are not correctly binarized.
For example,
the two regions marked by the blue circles
almost completely fall in the black phase;
but in the binarized slices filtered by NL-means,
a significant proportion of the two regions is still white.
We then conclude that {\sc PerSplat} filtering (e.g.,
% {\sc PerSplat-V$^{**}$})
{\sc PerSplat-V} with $\specsize=14659$)
provides the best binarized slices visually.
%\sout{with the above mentioned problems considerably mitigated}:
On one hand, the important details inside the white phase marked by the red circle
are preserved;
on the other hand, a larger proportion is classified as black in the two regions
marked by the blue circles.

\section{Methods}\label{sec:method}

\paragraph{Software used.}
MATLAB (R2019) was used to generate synthetic images, calculate binarization accuracy, and calculate tortuosity. 
ImageJ \cite{schindelin2012fiji} was used for an implementation of {\sc Otsu}.
{The Python library {\tt scikit-image}~\cite{van2014scikit} was used for implementations of
TV and NL-means.}

\paragraph{XCT Data used.}
One of the microstructures used in this study is a porous polymer, Polydimethylsiloxane (PDMS). 
This sample was made by AVP in DPB's lab using the process of templating where sugar cubes (Domino brand) were used as a template. The second sample is raw tomographic data obtained from~\cite{Pietsch2017} which is open-source. 
We performed X-ray Computed Tomography (XCT) on the porous PDMS. 
The sample in the open-source tomographic data is of commercially available lithium-ion batteries graphite electrode \cite{Muller2018}.  

%The anode from the {\red open source} is infiltrated by an epoxy to keep the microstructure intact during samples preparation and analysis. 
%\sout{With the anode being a carbon-based material and epoxy in the porosity,  it is hard to create contrast between different regions which creates significant errors in the image segmentation step. Binarization of these grey-scale images become difficult as peaks in the histograms tend to have more overlaps. In other words, not finding a right threshold value would greatly skew the results as significant data in the images will be wrongfully labelled. To binarize the data we used ImageJ and Otsu algorithm. As mentioned earlier that inspecting binarized data is very difficult as there is no notion of a ground truth. The next closest thing to the ground truth will be visualization of binary data and corresponding grey-scale data. But visualizing gigabytes of data will be cumbersome and could be highly partial.}\tao{This seems redundant.}

%\sout{Without the ground truth the algorithm we prepared to aid binarization was hard to test. So, we took grey-scale data of PDMS sample and anode sample, binarized using the Otsu algorithm. Using these same binary images, we then simulated images with similar grey-scale statistics but rooted in the same ground truth so that we could synthetically adjust the amount of overlap and the ability for our algorithm to segment and de-speckle identifiable phases in comparison to a known ground truth.}

\paragraph{Adding synthetic noise.}
To add noise to a given ground-truth image
in the process of synthetic data generation,
we took three steps. 
In step one, for each voxel from the black phase, 
we assign the voxel a grey-scale value based on a Gaussian distribution (for the black phase) of selected mean and variance.
For voxels from the white phase, we perform similar operations.
In step two, we used MATLAB's in-build function `{\sc{imnoise}}' to create Poisson noise. This MATLAB function's output depends on the type of input image (eg. 8 bit or 16 bit; our input was a 8 bit image).
As the last step, we do a simple nearest neighbor averaging of the grey-scale values.  This pays attention to near-neighbor voxel value averaging that often happens at boundaries between two phases in real XCT deconvolutions – as evidenced by the broad and flat region between the two Gaussians (see Figure~\ref{fig:1}).

\paragraph{X-ray tomography.}
{XCT images for porous polymer (PDMS) were produced using X-ray nano-computed tomography. Computed tomography is a non-destructive technique that allows full 3D spatial density maps of an object. To gather these tomographic images, two major steps were taken. First, the sample was inserted inside a CT machine called `SkyScan1272' manufactured by Bruker (Billerica, MA). A Hamamatsu L10101 micro-focus X-ray source was used with no filter. The X-ray source voltage and current was set to 40 kV and 200 $\mu\text{A}$, respectively, to get the best scan resolution. After setting up the X-ray source, a flat field correction was updated to minimize ring artifacts. A total scan of the sample contained 1472 projection images with a length and width of 2036 pixels by 2036 pixels, respectively. Images were taken at different angles equally spaced at 0.2$^{\circ }$ on a scale of 0$^{\circ }$ to 180$^{\circ }$, with exposure time of 225 ms and frame averaging of 5 per rotation. Secondly, to get 3D volume data, reconstruction of the above acquired raw data was done using a program called NRecon (Version: 1.7.4.6). The final resolution of each image was 4.50 $\mu\text{m}$/pixel.}

\section{Conclusion}
The investigation in this paper starts with an examination of the grey-scale histograms of commercial lithium-ion battery and a porous PDMS templated material.
%\sout{Our aim is to see how different types of histograms affect the ability of {\sc Otsu} algorithm to binarize.} \tamal{I feel this sentence is conveying that this is the main goal of the paper, which is not correct. I suggest deleting it.}
From the binarization results,
we find that  some important features of the microstructure are misinterpreted due to the histogram overlap, 
which led to miscalculation of microstructure characteristics. 
To solve the problem,
we design a novel filtering algorithm called {\sc {\sc PerSplat}} based on topological persistence 
to rectify the small regions that are assigned to wrong group otherwise during binarization (specifically by {\sc Otsu}). 
A tortuosity analysis and \emph{Minkowski functional} calculations (volume and surface area) show that the ground-truth values are better approximated when the data is preprocessed with {\sc PerSplat} and then binarized, as compared to a direct binarization.
Also, {\sc PerSplat} outperforms other filtering algorithms such as TV and NL-means in most cases.
%\sout{{\anand{and other filtering algorithms such as TV and NL-means}}.} 
The {\sc PerSplat} algorithm
has two variations when being applied on
3D data: {\sc PerSplat-P} and {\sc PerSplat-V}.
From our analysis, we observe that {\sc PerSplat-V} has more stability over various speckle sizes. 
We hope that this algorithm, in general, can make the microstructure characteristics computations more reliable and  hence help design better electrochemical devices. Additionally, the use of this algorithm is not only limited to battery research but potentially also to other applications engaging image analysis, ranging from tumor detection in biological sciences to detecting water content in geology. 

% \section{Code Availability}
% Requests for codes that support the findings within this paper should be addressed to T.K.D ({\tt tamaldey@purdue.edu}) and T.H ({\tt hou145@purdue.edu}) upon reasonable request.

% \section{Data Availability}
% The starting 3D grey-scale XCT microstructure data files are provided in Supplemental Materials.
% The X-ray computed tomography data used for the analysis of porous templated polymer is available online: \url{https://github.com/monk2k20/XCT-Polymer-Data}. The rest of the data that support the findings of this study are available from the corresponding author upon request.

%{\it ``Data availability statements should provide a statement about the availability of the minimal dataset that would be necessary to interpret, replicate and build upon the methods or findings reported in the article. Data availability statements should be provided as a separate section after the Methods section before the References, under the heading "Data Availability". For further guidance, please refer to the data availability and data citations policy information and Frequently Asked Questions (FAQs).''}

\section*{Acknowledgments}
The authors are grateful for support under NSF 1839267, NSF 1839252, NSF 2049010 as well as from the Rutgers Corning/Saint-Gobain/McLaren Endowment and Purdue Start-up fund. 

% {\red We would like to thank Yu-Min Chung and Sarah Day for providing the code implementation of PDThresholding.}

%A.V.P. fabricated the templated PDMS sample and performed XCT imaging. T.H. and T.K.D. designed the {\sc PerSplat} algorithm and T.H. implemented the algorithm. A.V.P. and J.D.B. wrote the synthetic image generation code and the code for calculating binarization accuracy as well as performing tortuosity calculations. D.P.B. and T.K.D. provided combined supervision and guidance throughout. A.V.P. and T.H. wrote the manuscript. All authors edited the paper and approved its final version.

% \newpage
% \section{Competing Interests}
% The authors declare no competing interests.

% \newpage

% \bibliographystyle{plain}
\bibliographystyle{unsrt}
\bibliography{refs}

%ref 10,26

\end{document}